\documentclass{pasj00}

\def\commenta{$^*$}
\def\commentb{$^\dagger$}
\def\commentc{$^\ddagger$}
\def\commentd{$^\S$}
\def\commente{$^\|$}

\SetRunningHead{Y. Osaki and T. Kato}{Superoutbursts 
and Superhumps in SU UMa Stars}

\title{A Further Study of Superoutbursts and Superhumps in SU UMa Stars
by the Kepler Light Curves of V1504 Cygni and V344 Lyrae}

\author{Yoji \textsc{Osaki}}
\affil{Department of Astronomy, School of Science, University of Tokyo,
Hongo, Tokyo 113-0033}
\email{osaki@ruby.ocn.ne.jp}

\and

\author{Taichi \textsc{Kato}}
\affil{Department of Astronomy, Kyoto University,
       Sakyo-ku, Kyoto 606-8502}
\email{tkato@kusastro.kyoto-u.ac.jp}


\begin{document}

\maketitle

\begin{abstract}
We made a supplemental study of the superoutbursts and superhumps in 
SU UMa stars by using the recently released Kepler public data 
of V1504 Cyg and V344 Lyr. 
One of the superoutbursts in V1504 Cyg was preceded 
by a precursor normal outburst which was well separated from 
the main superoutburst. The superhump first appeared 
during the descending branch of the precursor normal outburst 
and it continued into quiescence (the deep dip between the precursor 
and the main superoutburst), and it began to grow in amplitude 
with the growth of the main superoutburst after quiescence ended. 
A similar phenomenon was also observed in V344 Lyr. 
This observation demonstrates very clearly that the superoutburst was 
triggered by the superhump (i.e., by the tidal instability), supporting 
the thermal-tidal instability model. 
Smak (2013, Acta Astron., 63, 109) criticized our previous paper 
(Osaki and Kato, 2013, PASJ, 65, 50) and challenged our main 
conclusion that various observational lines of evidence of V1504 Cyg 
support the thermal-tidal instability model for the superoutbursts of 
SU UMa stars.  We present our detailed accounts to all of his criticisms 
by offering clear explanations. 
We conclude that the thermal-tidal 
instability model is after all only the viable model for the superoutbursts 
and superhumps in SU UMa stars. 
\end{abstract}

\section{Introduction}

The SU UMa stars are one of the sub-classes of dwarf novae characterized 
by the so-called ``superoutburst'' and ``superhumps''. It is now well 
established that the outbursts of dwarf novae are produced 
by a sudden brightening of the accretion disk around the central white dwarf 
in the semi-detached close binary systems in which the red-dwarf 
secondary star supplies mass to the accretion disk.
In addition to the ordinary normal outbursts observed in the 
U Gem type dwarf novae, the SU UMa stars show a longer and brighter outburst 
called the superoutburst. During a superoutburst, a periodic photometric 
hump with amplitude 0.2--0.3 mag appears and it is called the superhump 
whose period is very near to the orbital period of the system 
but it is longer than that by a few percent. 
The ordinary superhump mentioned above is also called the ``positive" 
superhump (abbreviated as pSH) because another periodic hump 
with a period shorter than the 
orbital period appears in some SU UMa and nova-like variable stars and the 
latter is called the ``negative'' superhump (nSH). 
The origin of the positive superhump 
is now understood as due to a deformation of accretion disk into eccentric 
form and a slow apsidal precession of the eccentric disk is responsible 
for the basic clock of the positive superhump. On the other hand the negative 
superhump is now thought to be produced by a tilted accretion disk whose 
nodal line precesses retrograde, giving rise to the underlying clock of the 
negative superhump. The general reviews on these subjects are 
found in \citet{war95book} and \citet{hel01book}. 

In a series of papers we have studied the superoutbursts and superhumps 
in SU UMa stars by using the Kepler light curves of SU UMa stars.  
In the first paper \citep{osa13v1504cygKepler}
(hereafter referred to Paper I) 
we have studied the Kepler light curve of an SU UMa star, V1504 Cyg, 
and we have demonstrated that various observational lines of evidence 
in V1504 Cyg support the thermal-tidal instability (TTI) model 
for the superoutburst of SU UMa stars proposed by \citet{osa89suuma} 
(for a review on the TTI model, see, \cite{osa96review}).
In particular, we have shown that the frequency of the negative superhump 
varies systematically during a supercycle 
(a cycle from one superoutburst to the next) and by interpreting 
those variations basically as the result of variations in 
the disk radius, we have demonstrated that observed disk radius variation 
fits very well with a prediction of the thermal-tidal instability model. 
In the second paper \citep{osa13v344lyrv1504cyg} (Paper II),
we have examined the Kepler short-cadence light curves of two SU UMa stars, 
V344 Lyr and V1504 Cyg.  
We have analyzed the simultaneous frequency variations of the positive 
and negative superhumps.  We have demonstrated that these 
two signals vary in unison, if appropriately converted, 
indicating the disk radius variation. 

The Kepler public data are released every three-months and the new data  
continue to arrive. As shown below, 
the newly released data of V1504 Cyg and V344 Lyr have basically supported  
our main conclusions reached in our Paper I and Paper II. 
However, the new data also give us some surprise and in 
section \ref{sec:analysisnewdata} 
we made a supplemental study to our Paper I and Paper II 
on the superoutbursts and superhumps of V1504 Cyg and V344 Lyr 
by using newly released public Kepler data. 

\citet{sma13negSH} has recently criticized our Paper I by arguing that our 
interpretation and our main conclusion about the nature of superoutbursts 
were incorrect.  As his preprint already appeared in astro-ph 
(arXiv:1301.0187) on 2013 January 2, we examined in our Paper II  
his criticism about our interpretation on variation of 
the negative superhump periods. 
Since his paper has finally appeared in a published form in 
Acta Astronomica, here we present our own accounts once again 
to almost of all his criticisms in section \ref{sec:replytoSmak}. 

This paper consists of two parts: the first part concerns 
a supplemental study of newly released Kepler data of V1504 Cyg and 
V344 Lyr in section \ref{sec:analysisnewdata}, 
and the second part our reply to 
Smak's criticisms in section \ref{sec:replytoSmak}.
A summary is given in section \ref{sec:summary}. 

\section{A Further Study of the Superoutburst and Superhumps by Using 
Newly Released Kepler Light Curves of V1504 Cyg 
and V344 Lyr} \label{sec:analysisnewdata}

When we wrote our Paper II in March 2013, the Kepler data available 
for the public were those from the quarters 2 to 10 (Q2--10) and 
the quarter 14 (Q14).  Since then, data for two more quarters 
Q11 and Q15 were released in April 2013 and then further data for 
two more quarters, Q12 and Q16, released in July 2013. 
In total, the short-cadence 
Kepler Data for V1504 Cyg and V344 Lyr available to the public now 
extend from June 2009 to April 2013, well over three years from Barycentric 
Julian Date (BJD) 2455002 to 2456390 except for the quarter 13 (Q13), 
which will be releasd in October 2013. 
The photometric observations by the Kepler 
satellite were unfortunately stopped in May 2013 because of problems  
in the reaction wheels onboard the satellite.  Its recovery is still 
uncertain. The data from the quarter 16 could be the last.

In this paper, we examine those new data of V1504 Cyg and V344 Lyr 
for Q10--12 and Q14--16. Analysis of those for Q2--Q10 has been presented 
in Paper II, which should be consulted whenever it is necessary. 
We first summarize global light curves and the power spectra 
for newly released data of V1504 Cyg and V344 Lyr 
in subsection \ref{sec:globalLC}.
We then discuss three problems in this section. 
We found that one of superoutbursts in V1504 Cyg  
was preceded by a precursor normal outburst which was 
well separated from the main superoutburst. 
The same type of precursor was also found in V344 Lyr and 
we study them in subsection \ref{sec:precNO}. 
We then examine the failed superhump which occurred in the descending 
branch of one of normal outbursts just prior to a superoutburst 
but failed to trigger a superoutburst in subsection \ref{sec:failedSH}. 
In subsection \ref{sec:miniNO} we discuss  
mini-outbursts which occurred exclusively in the later half of 
the Type S supercycle of V1504 Cyg, a supercycle not accompanied by 
negative superhumps. 

\subsection{Global Light Curves and the Power Spectra for the Newly Released  
Kepler data of V1504 Cyg and V344 Lyr} \label{sec:globalLC}

\begin{figure*}
  \begin{center}
%  \FigureFile(160mm,120mm){v1504spec2dseq8.eps}
  \FigureFile(160mm,120mm){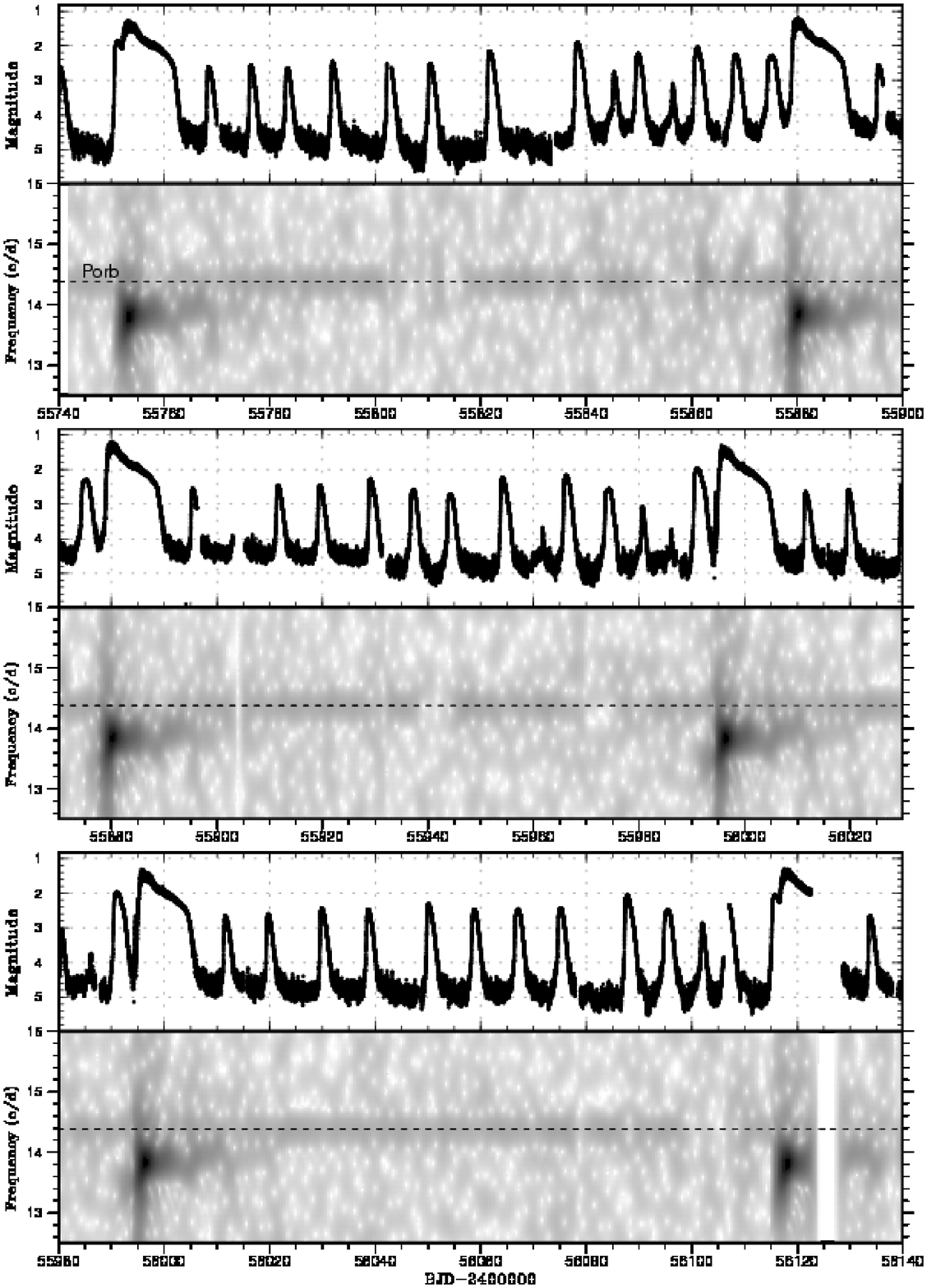}
  \end{center}
  \caption{Two-dimensional discrete Fourier power spectrum 
  of the Kepler light curve of V1504 Cyg for supercycles 8--10.
  From the top to the bottom for each supercycle, 
  Upper: light curve; the Kepler data were binned to 0.005~d,
  Lower: power spectrum. 
  The sliding window and the time step used are 5~d and 0.5~d,
  respectively.}
  \label{fig:v1504spec2d}
\end{figure*}

First we made two-dimensional discrete Fourier transform (FT)
power spectra for the newly released data of quarters Q10--12 
and Q14--16 of V1504 Cyg and V344 Lyr, and they are shown in 
figures \ref{fig:v1504spec2d} and \ref{fig:v344spec2d}\footnote{
  As mentioned in note added in proof, the original figures 1--4 were 
  replaced by new ones, by including data for Q13.
}.
The method of analysis was the same as in described in Paper I.
The two-dimensional least absolute shrinkage and selection operator
(Lasso) power spectra (\cite{kat12perlasso}; \cite{kat13j1924})
are also shown in figures \ref{fig:v1504spec2dlasso} and
\ref{fig:v344spec2dlasso} as in paper II.

We summarize the main characteristics of supercycles and superoutbursts 
in table \ref{tab:supercycleV1504} for V1504 Cyg and table 
\ref{tab:supercycleV344} for V344 Lyr as in Paper I and Paper II \footnote{
As mentioned in note added in proof, the original tables 1-2 were 
replaced by new ones, by including data for Q13.
}. 
The symbols used in these two tables are the same as those given there 
and no explanations may be needed. 

\addtocounter{figure}{-1}
\begin{figure*}
  \begin{center}
%  \FigureFile(160mm,120mm){v1504spec2dseq11.eps}
  \FigureFile(160mm,120mm){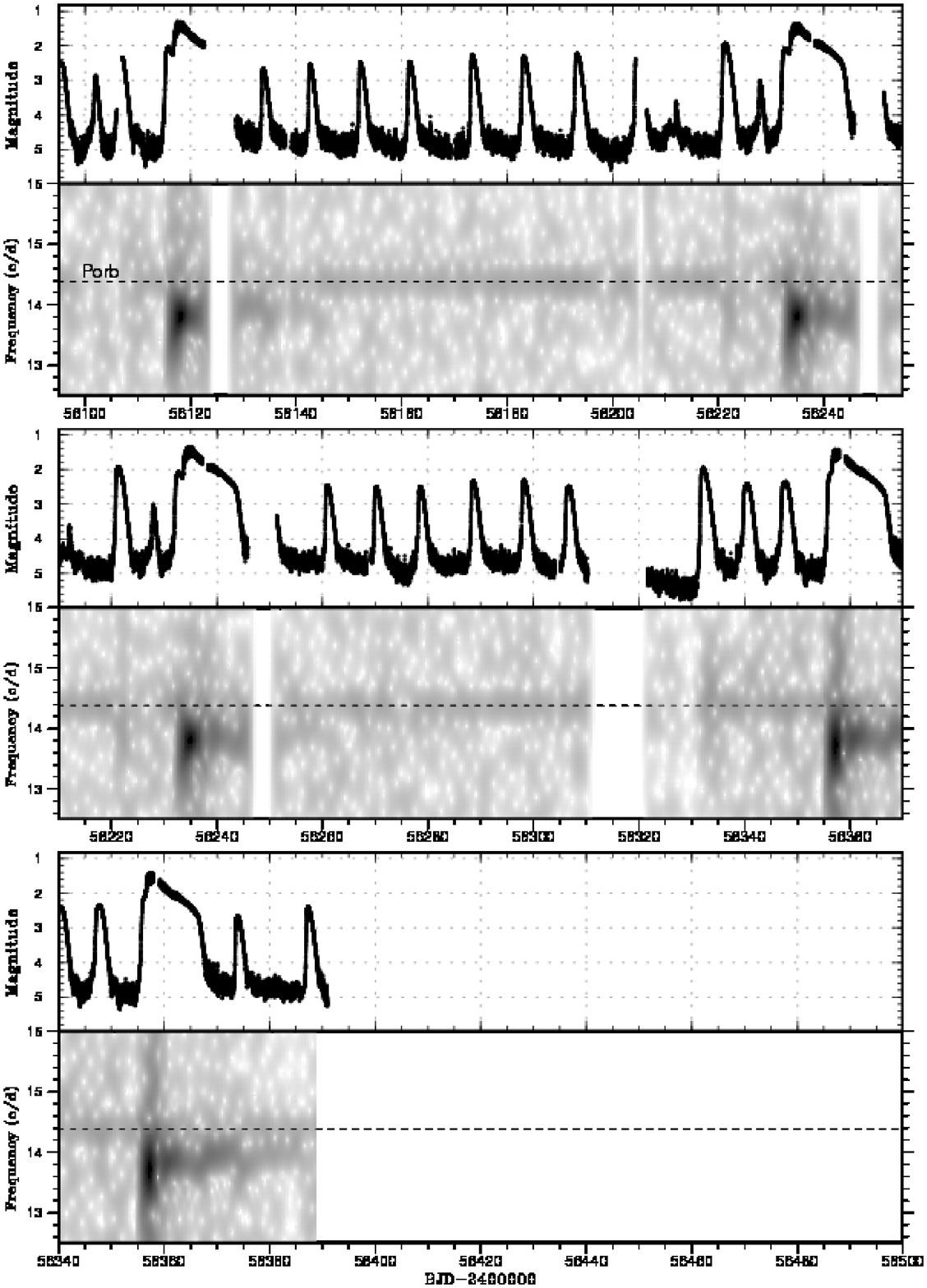}
  \end{center}
  \caption{Two-dimensional discrete Fourier power spectrum 
  of the Kepler light curve of 
  V1504 Cyg for supercycles 11--13.
  From the top to the bottom for each supercycle,  
  Upper: light curve; the Kepler data were binned to 0.005~d,
  Lower: power spectrum. 
  The sliding window and the time step used are 5~d and 0.5~d,
  respectively.}
\end{figure*}

\begin{figure*}
  \begin{center}
%  \FigureFile(160mm,120mm){v344spec2dseq8.eps}
  \FigureFile(160mm,120mm){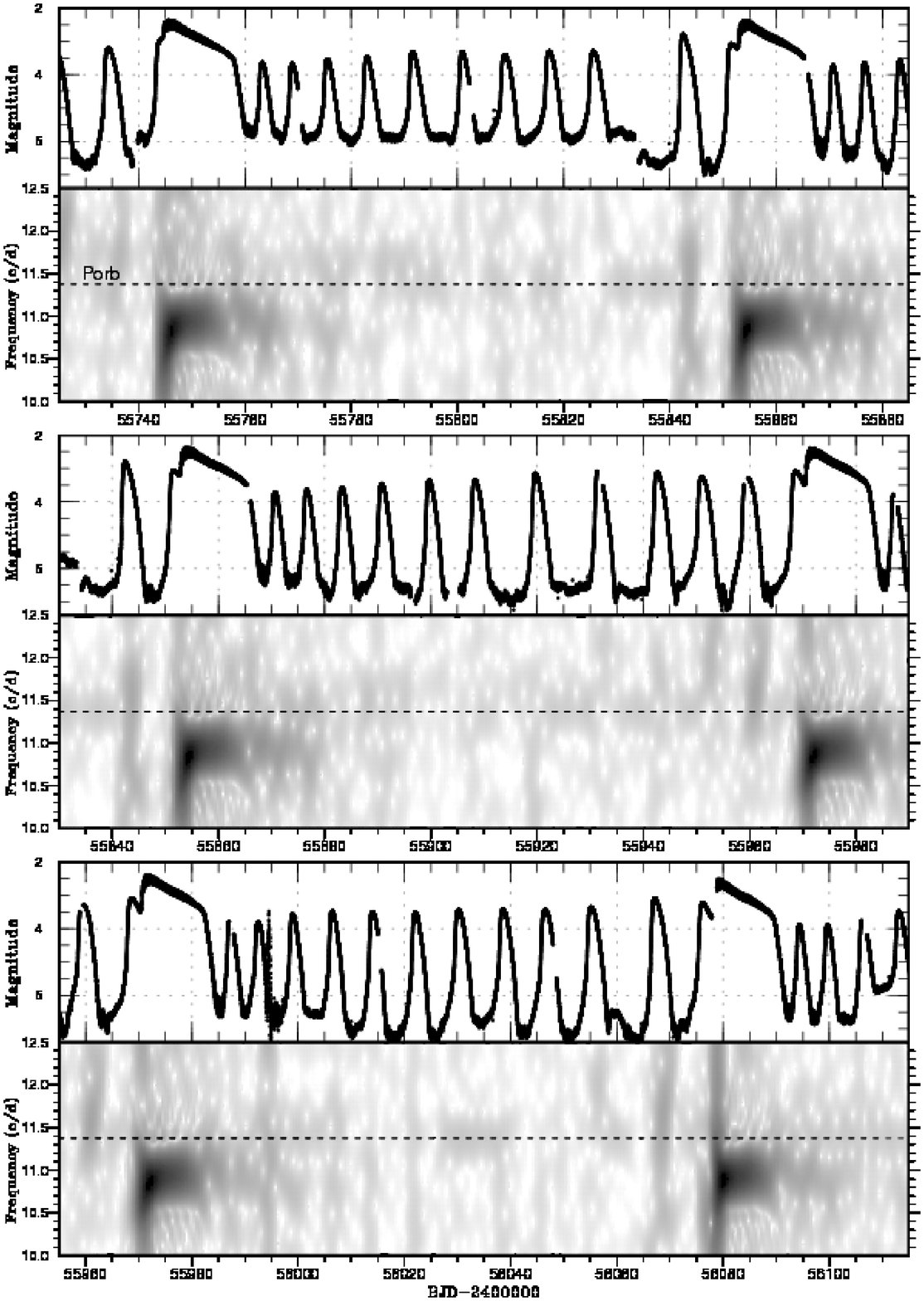}
  \end{center}
  \caption{Two-dimensional discrete Fourier power spectrum 
  of the Kepler light curve of V344 Lyr for supercycles 8--10.
  From the top to the bottom for each supercycle, 
  Upper: light curve; the Kepler data were binned to 0.005~d,
  Lower: power spectrum
  the sliding window and the time step used are 5~d and 0.5~d,
  respectively.}
  \label{fig:v344spec2d}
\end{figure*}
\addtocounter{figure}{-1}
\begin{figure*}
  \begin{center}
%  \FigureFile(160mm,120mm){v344spec2dseq11.eps}
  \FigureFile(160mm,120mm){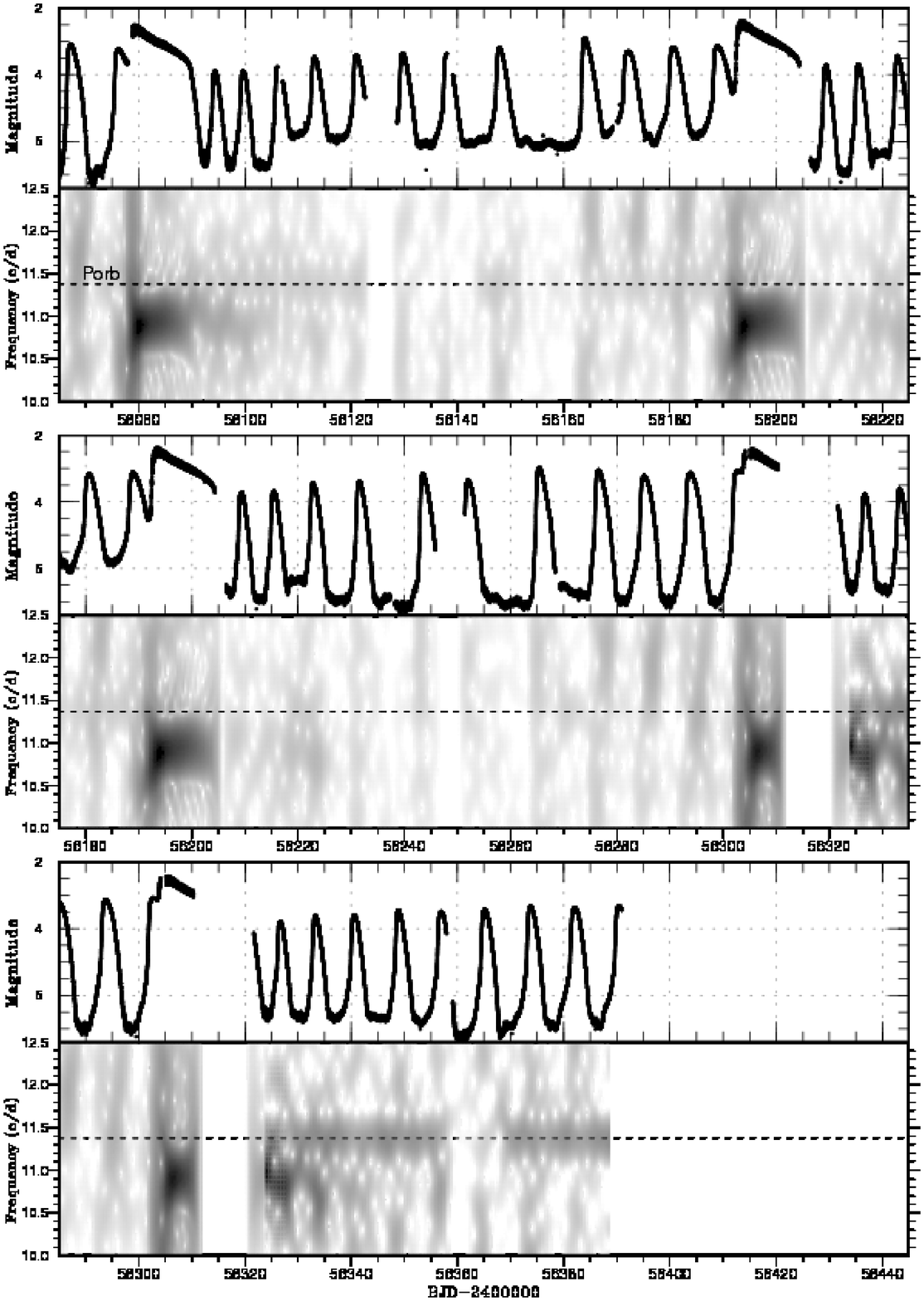}
  \end{center}
  \caption{Two-dimensional discrete Fourier power spectrum 
  of the Kepler light curve of 
  V344 Lyr for supercycles 11--13.
  From the top to the bottom for each supercycle,  
  Upper: light curve; the Kepler data were binned to 0.005~d,
  Lower: power spectrum. 
  The sliding window and the time step used are 5~d and 0.5~d,
  respectively.}
\end{figure*}

\begin{figure*}
  \begin{center}
%  \FigureFile(160mm,120mm){v1504spec2dlasso8.eps}
  \FigureFile(160mm,120mm){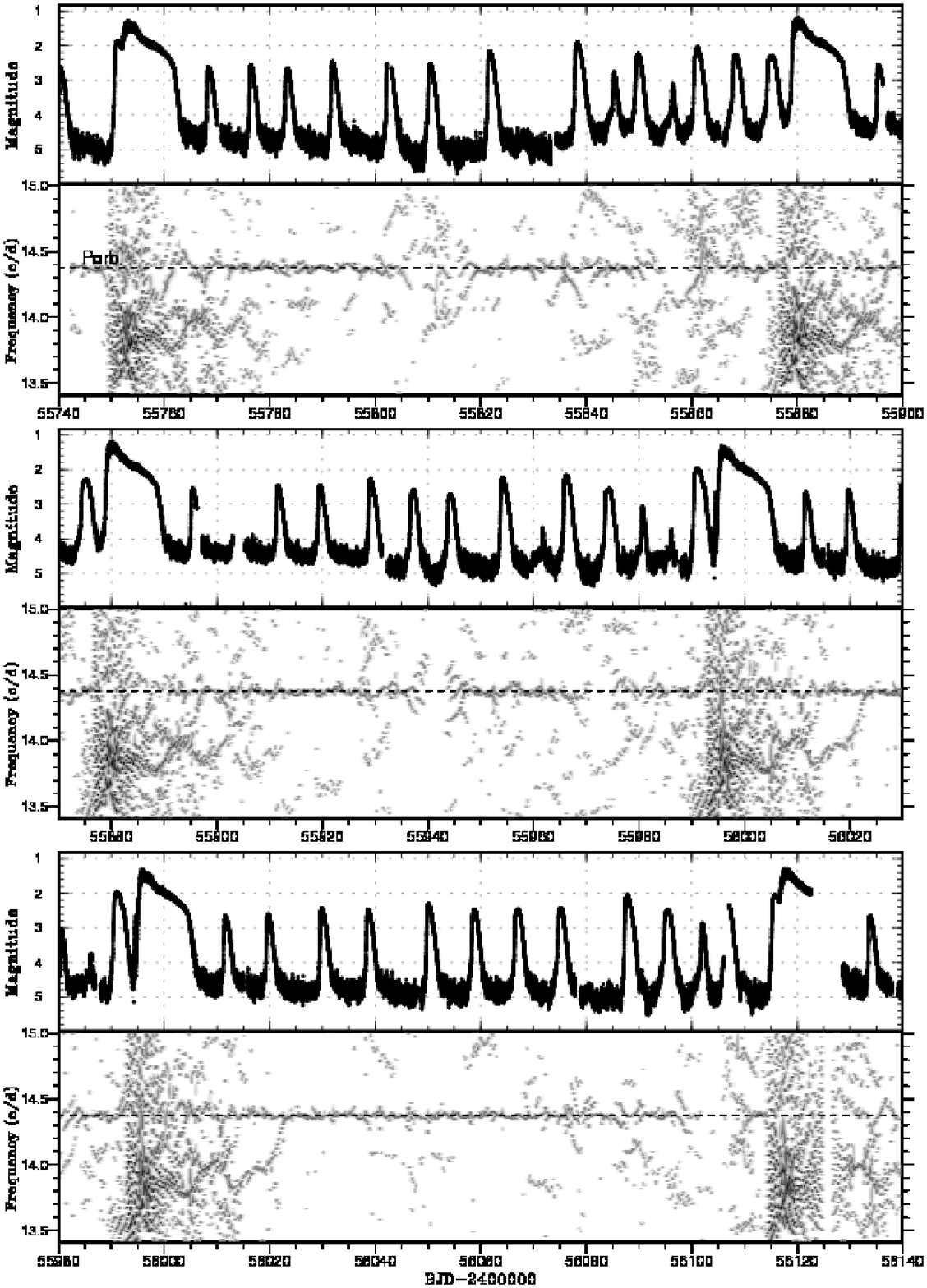}
  \end{center}
  \caption{Two-dimensional lasso power spectrum 
  of the Kepler light curve of V1504 Cyg for supercycles 8--10.
  From the top to the bottom for each supercycle, 
  Upper: light curve; the Kepler data were binned to 0.005~d,
  Lower: power spectrum
  the sliding window and the time step used are 5~d and 0.5~d,
  respectively.}
  \label{fig:v1504spec2dlasso}
\end{figure*}
\addtocounter{figure}{-1}
\begin{figure*}
  \begin{center}
%  \FigureFile(160mm,120mm){v1504spec2dlasso11.eps}
  \FigureFile(160mm,120mm){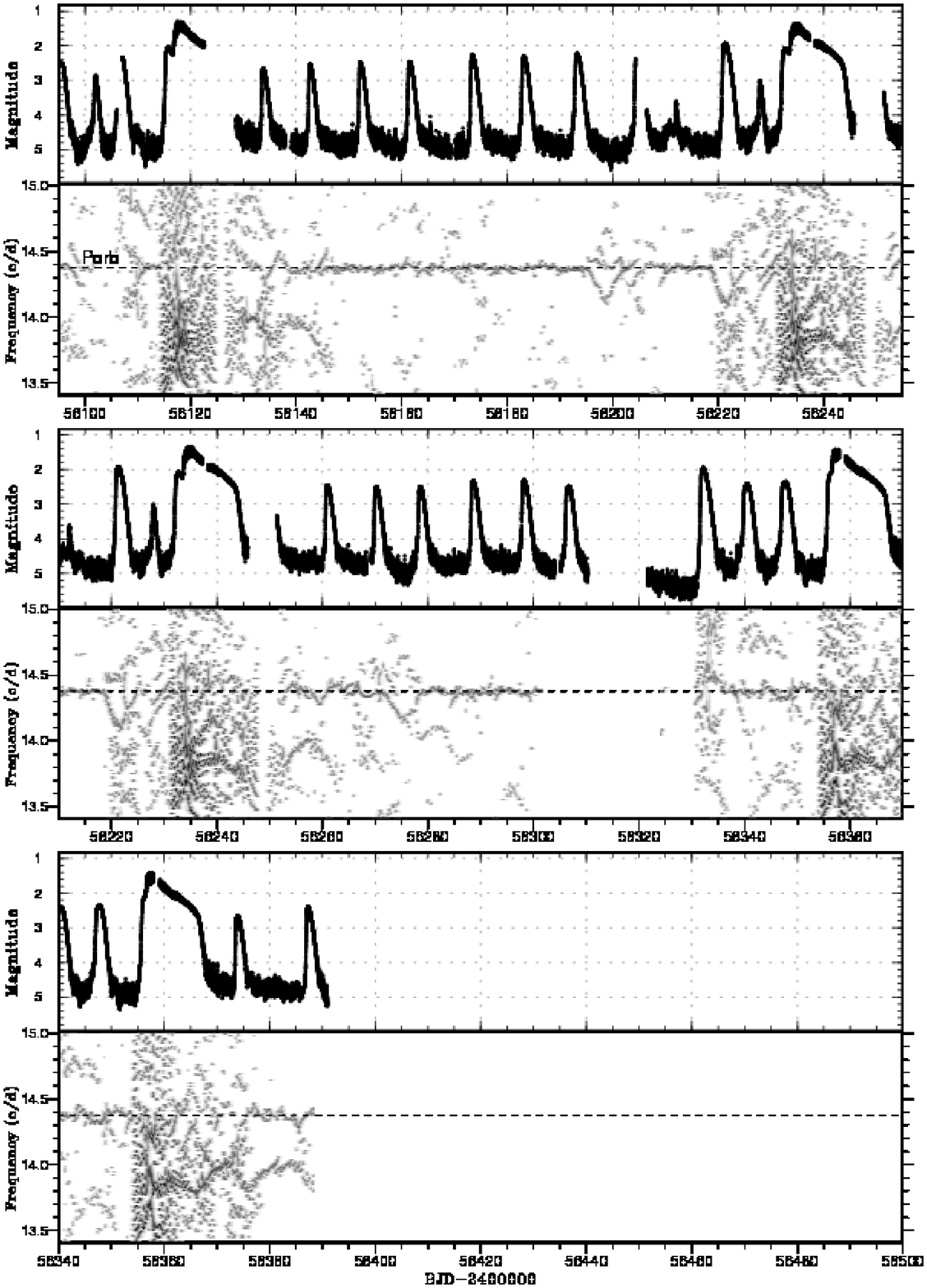}
  \end{center}
  \caption{Two-dimensional lasso power spectrum 
  of the Kepler light curve of 
  V1504 Cyg for supercycles 11--13.
  From the top to the bottom for each supercycle,  
  Upper: light curve; the Kepler data were binned to 0.005~d,
  Lower: power spectrum. 
  The sliding window and the time step used are 5~d and 0.5~d,
  respectively.}
\end{figure*}

\begin{figure*}
  \begin{center}
%  \FigureFile(160mm,120mm){v344spec2dlasso8.eps}
  \FigureFile(160mm,120mm){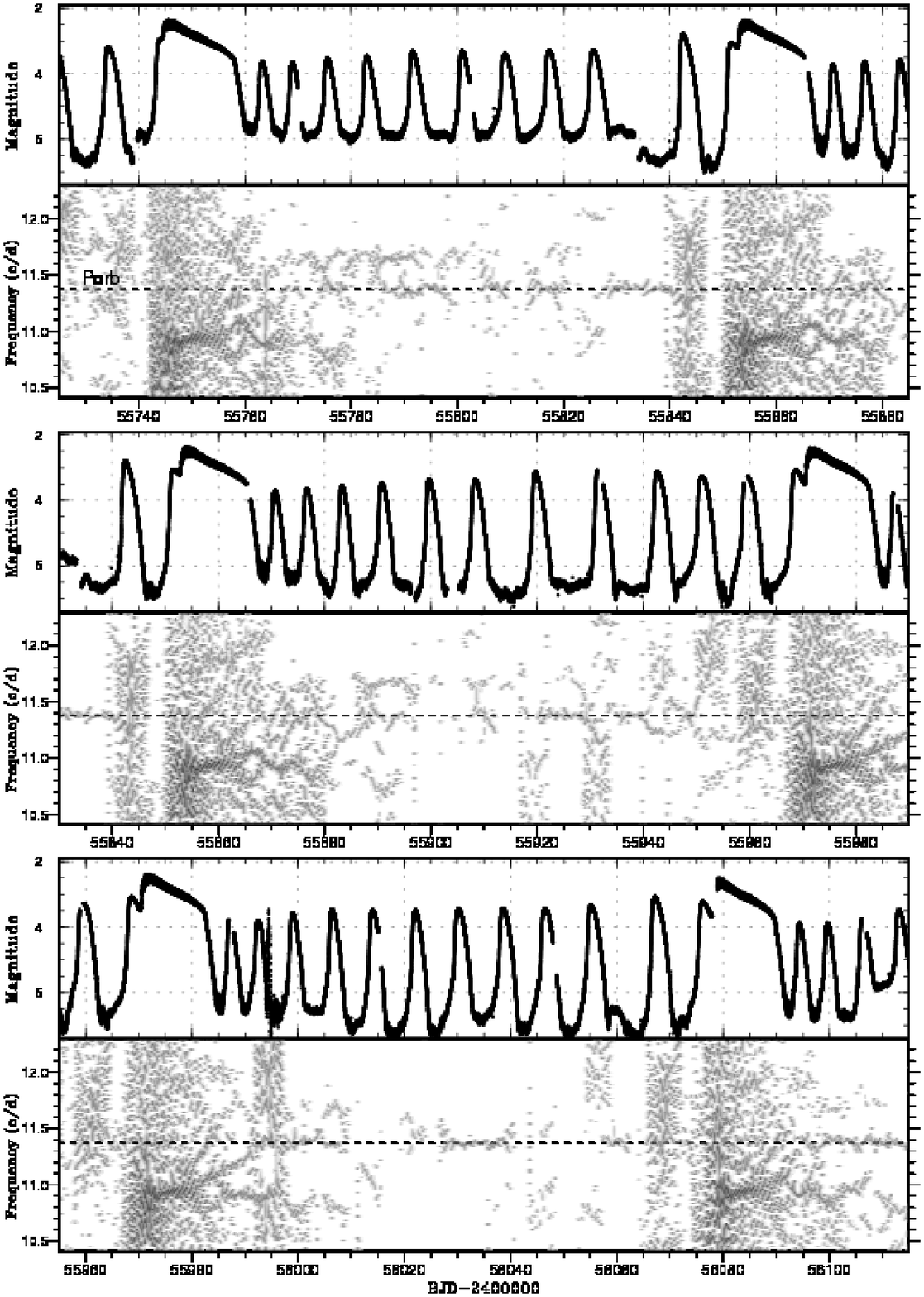}
  \end{center}
  \caption{Two-dimensional Lasso  power spectrum 
  of the Kepler light curve of V344 Lyr for supercycles 8-10.
  From the top to the bottom for each supercycle, 
  Upper: light curve; the Kepler data were binned to 0.005~d,
  Lower: power spectrum. 
  The sliding window and the time step used are 5~d and 0.5~d,
  respectively.}
  \label{fig:v344spec2dlasso}
\end{figure*}
\addtocounter{figure}{-1}
\begin{figure*}
  \begin{center}
%  \FigureFile(160mm,120mm){v344spec2dlasso11.eps}
  \FigureFile(160mm,120mm){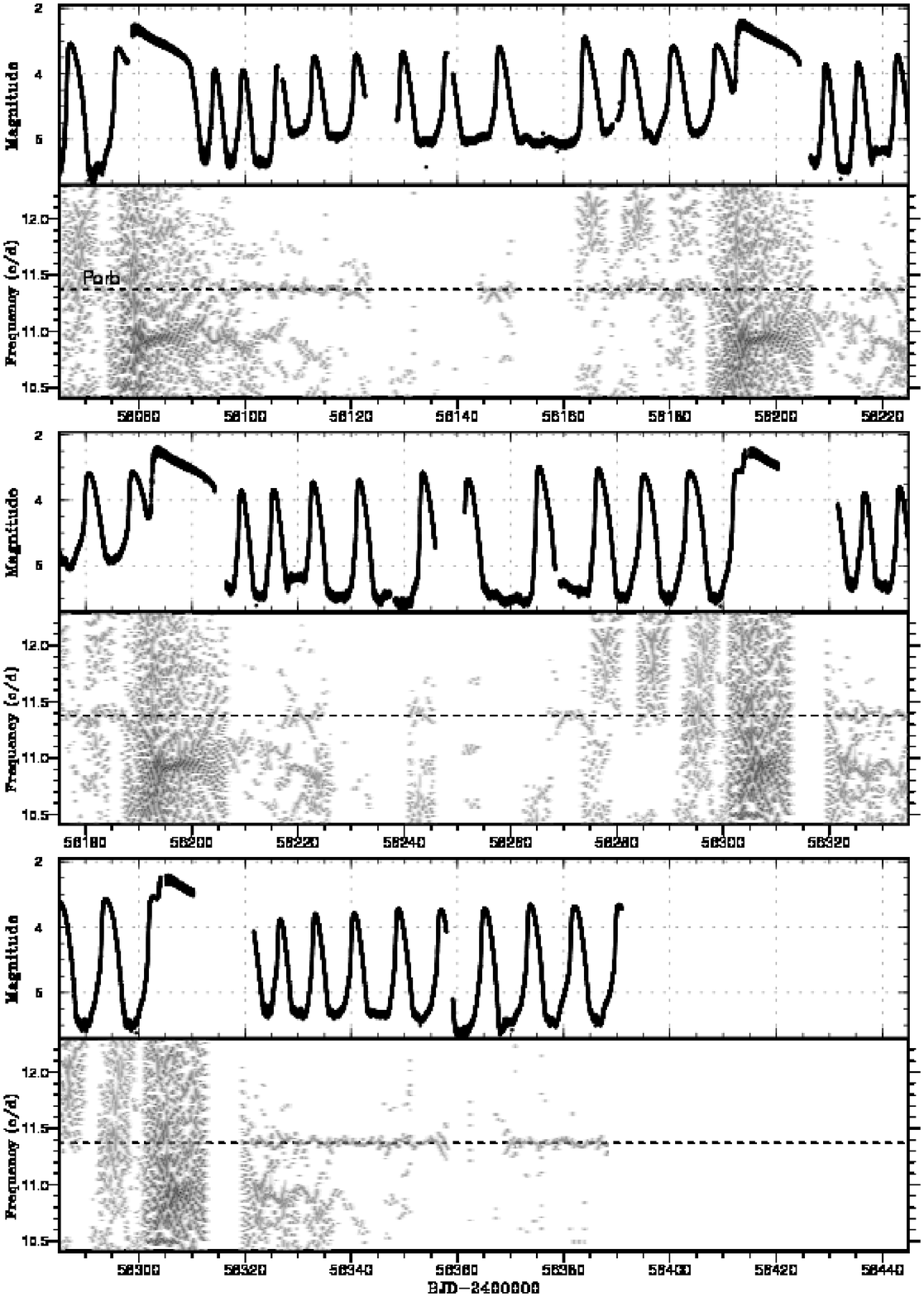}
  \end{center}
  \caption{Two-dimensional Lasso power spectrum 
  of the Kepler light curve of 
  V344 Lyr for supercycles 11--13.
  From the top to the bottom for each supercycle,  
  Upper: light curve; the Kepler data were binned to 0.005~d,
  Lower: power spectrum. 
  The sliding window and the time step used are 5~d and 0.5~d,
  respectively.}
\end{figure*}

\begin{table*}
\caption{Superoutbursts and supercycles of V1504 Cyg. \commenta}
        \label{tab:supercycleV1504}
\begin{center}
\begin{tabular}{cccccccccc}
\hline
(1) & (2) & (3) & (4) & (5) & (6) & (7) & (8) & (9) & (10) \\
SC & start & start & end & SC length  & SO & SC & number & negative & orbital \\
number & of SC\commentb & of SO\commentb & of SO\commentb & excluding SO\commentc & duration\commentc & length\commentc & of NO      &   SH         & hump \\
\hline
1 &  --  & 74.5  & 88.5 &  --   & 14   & $>$88 & $>$8 & no & no \\
2 & 88.5 & 201   & 215  & 112.5 & 14   & 126.5 &  10  & no & no \\
3 & 215  & 312   & 325  &  97   & 13   & 110   &  9  & later half & partly \\
4 & 325  & 406.5 & 419  &  81.5 & 12.5 &  94   &   6  & full & visible \\
5 & 419  & 517   & 530  &  98   & 13   & 111   &   5  & full & partly \\
6 & 530  & 638\commentd  & 650 & 108
& 12  & 120  & 10  & early part & later part \\
7 & 650 & 750   & 763.5  & 100 & 13.5 & 113.5 &   12  & no & yes \\
8 & 763.5  & 878 & 890  &  114.5   &  12  &  126.5 & 14\commente & no & yes \\
9 & 890 & 994.5 & 1006 & 104.5  &  11.5 & 116  & 14\commente & no & yes \\
10 & 1006 & 1115   & 1128  & 109 & 13 & 122 & 12  & no & yes \\
11 & 1128 & 1232   & 1245  & 104 & 13 & 117 &   11  & no & yes \\
12 & 1245 & 1355   & 1368  & 110 & 13 & 123 &   $\ge$ 10  & no & yes \\
13 & 1368  &  --    &  --   & --  & -- & --  & --&  -- & -- \\
\hline
  \multicolumn{10}{l}{\commenta Abbreviations in this table: supercycle (SC), superoutburst (SO), normal outburst (NO), superhump (SH).} \\
  \multicolumn{10}{l}{\commentb BJD$-$2455000.} \\
  \multicolumn{10}{l}{\commentc Unit: d.} \\
  \multicolumn{10}{l}{\commentd date guessed because of data gap} \\
  \multicolumn{10}{l}{\commente precursor normal outburst included} \\
\end{tabular}
\end{center}
\end{table*}

\begin{table*}
\caption{Superoutbursts and supercycles of V344 Lyr.\commenta}
        \label{tab:supercycleV344}
\begin{center}
\begin{tabular}{cccccccccc}
\hline
(1) & (2) & (3) & (4) & (5) & (6) & (7) & (8) & (9) & (10) \\
SC & start & start & end & SC length  & SO & SC & number & negative & orbital \\
number & of SC\commentb & of SO\commentb & of SO\commentb & excluding SO\commentc & duration\commentc & length\commentc & of NO      &   SH         & hump \\
\hline
1 & -- & 55.5 & 73 & --& 17.5 & $>$70 & $>$4 & well visible  & no \\
2 & 73 & 160 & 178 & 87 & 18   & 105 &  7  & well visible & no \\
3 & 178  & 277   & 294  &  99  & 17 & 116   &  7  & no & visible \\
4 & 294  & 397 & 415  &  103 & 18 & 121   &   10  & no & visible \\
5 & 415 & 527   & 544.5 & 112   & 17.5   & 129.5 &  13  & no & visible  \\
6 & 544.5 & 641.5  & 659.5 & 97 &  18 & 115 & $\ge$10 & no & well visible \\
7 & 659.5 & 742   & 760  & 82.5 & 18 & 100.5 &   8  & well visible & no \\
8 & 760  & 850  & 868  &  90   &  18  & 108  &  10 &partly visible & partly 
visible  \\
9 & 868 & 966 & 985 & 98 & 19 & 117 & 11 & partly visible &partly visible \\
10 & 985  & 1074 & 1092  &  89 & 18 & 107   &  11  & no & partly visible \\
11 & 1092  & 1188 & 1206  & 96 & 18 & 114   &  11  & no & partly visible \\
12 & 1206  & 1301 & 1319\commentd &  95 & 18 & 113   &  10  & no & no \\
13 & 1319  & -- & --  &  -- & -- & --   & $\ge$  10  & no & yes \\
\hline
  \multicolumn{10}{l}{\commenta Abbreviations in this table: supercycle (SC), superoutburst (SO), normal outburst (NO), superhump (SH).} \\
  \multicolumn{10}{l}{\commentb BJD$-$2455000.} \\
  \multicolumn{10}{l}{\commentc Unit: d.} \\
   \multicolumn{10}{l}{\commentd date guessed because of data gap} \\
\end{tabular}
\end{center}
\end{table*}

Figures \ref{fig:v1504spec2d} and \ref{fig:v1504spec2dlasso} exhibit 
light curves and power spectra of V1504 Cyg for supercycles Nos. 8--13 and 
figures \ref{fig:v344spec2d} and \ref{fig:v344spec2dlasso} do 
those of V344 Lyr for supercycles Nos. 8--13.  
Most of supercycles in both V1504 Cyg and V344 Lyr shown here are 
the Type S supercycle (i.e., lacking the negative 
superhumps) except for the supercycle No. 8 of V344 Lyr in which 
the ordinary negative superhumps were visible 
in its early part.  The newly released Kepler data of the two stars 
basically confirm the results of Paper I and Paper II.
That is, a good correlation 
between the appearance of the negative superhumps and the quiescence 
interval of two consecutive outbursts (and thus the number of normal 
outbursts in a supercycle), which led us to the classification of supercycles: 
the Type L supercycle accompanied by the negative 
superhumps and the Type S supercycle without the negative superhumps. 

Let us examine more closely the outburst characteristics of V1504 Cyg which 
are summarized in table \ref{tab:supercycleV1504}. As seen in table 
\ref{tab:supercycleV1504}, the Kepler data used here 
now cover all 12 superoutbursts of V1504 Cyg 
and its supercycles 2-12 except for a supercycle 10 which 
occurred in Q13. As discussed in Paper I, the supercycles No.4 and No. 5 
of V1504 Cyg were well defined Type L supercycles in which a strong signal of 
the negative superhumps appeared. Supercycles No.8-9 and No.11-12 shown in 
figures \ref{fig:v1504spec2d} and \ref{fig:v1504spec2dlasso} are found 
to be the Type S supercycles without negative superhumps. The mean supercycle 
length of V1504 Cyg is found to be 116.3d. 
As seen in table \ref{tab:supercycleV1504}, the supercycle length is 
well correlated with the supercycle Types: short for the 
Type L supercycles and long for the Type S.  From this correlation, we can 
guess that the supercycle No. 10 must be the Type S supercycle because of 
a long supercycle length ($\sim 122$d), which can be checked in October 2013. 
Particularly interesting is a supercycle No. 8 of V1504 Cyg where 
14 normal outbursts occurred in a supercycle, including 
two mini-outbursts. This will be discussed in subsection \ref{sec:miniNO}. 

Let us now turn our attention to the case of V344 Lyr. The light curve and 
power spectra of newly released data of V344 Lyr are shown 
in figures \ref{fig:v344spec2d} and \ref{fig:v344spec2dlasso} and 
the characteristics of supercycles are summarized in table  
\ref{tab:supercycleV344}. As seen in these figures, a weak signal of the 
negative superhump (nSH) is visible in the early part of supercycle No. 8 
and also a much weaker signal is barely visible in the early part of 
supercycle No. 9. We also note from these figures 
that all superoutbursts of V344 Lyr were preceded by one to three 
normal outbursts accompanied by the impulsive negative superhump 
(which was found by \citet{woo11v344lyr} and discussed in Paper II).  
These observations basically confirm our conclusion reached in Paper II: 
the occurrence of the negative superhump in V344 Lyr was more patchy 
than the case of V1504 Lyr, and all superoutbursts of V344 Lyr 
were preceded by normal outbursts accompanied by the impulsive nSH.  

\subsection{A Precursor Normal Outburst Well Separated from the Main 
Superoutburst} \label{sec:precNO}

All superoutbursts in V1504 Cyg and V344 Lyr so far studied in Paper I 
and Paper II have turned out to be precursor-main types 
in which the precursor normal outburst merged quite well with the main 
superoutburst and it formed a part of superoutburst. 

\begin{figure*}
  \begin{center}
%    \FigureFile(140mm,190mm){v1504spec2dfreqso8.eps}
    \FigureFile(140mm,190mm){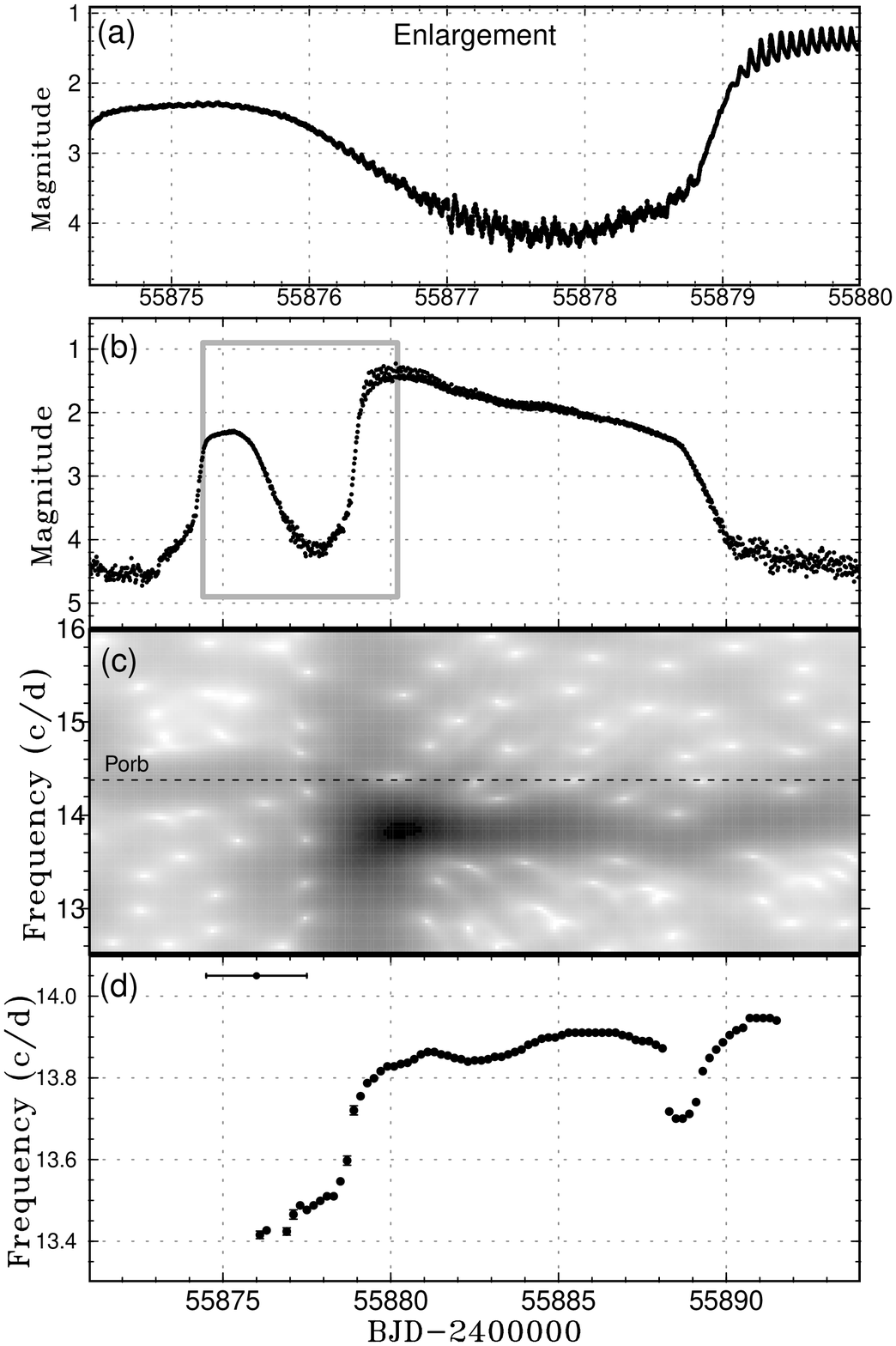}
  \end{center}
  \caption{
  Kepler light curve of superoutburst No. 8 of V1504 Cyg.
  (a) Enlargement of the box in panel b. The Kepler data were 
  binned to 0.001~d.  Superhumps developed
  between the short outburst and the superoutburst.
  (b) Light curve corresponding to the interval of panels c and d;
  the Kepler data were binned to 0.01~d
  (c) Two-dimensional power spectrum.
  The sliding window and the time step used are 5~d and 0.1~d,
  respectively.
  (d) Frequency variation of superhumps.
  The initial low frequency
  superhumps correspond to the growing stage superhumps (stage A
  superhumps).
  }
  \label{fig:v1504spec2dfreqso8}
\end{figure*}

However a new type of superoutburst occurred in superoutburst 
No. 8 at the day 55880 of V1504 Cyg (where the day is counted 
from BJD 2400000) in which a precursor normal outburst was 
separated by a deep dip (quiescence) from the main superoutburst so that 
the precursor normal outburst looked as an isolated normal outburst. 
Figure \ref{fig:v1504spec2dfreqso8} (panel b) shows 
this part of light curve of V1504 Cyg. 
A normal outburst which occurred at the day 55875 looked like 
an ordinary normal outburst; there is no clear difference from 
the preceding two normal outbursts in light curve.
The next superoutburst occurred  4~d after the maximum of this 
normal outburst.  At a first look, this superoutburst looked different 
from those studied before as it looked as if not accompanied by a precursor 
because the star rose straightly up to a light maximum from quiescence. 

In reality, what happened is that the preceding normal outburst at the day
55875 was a precursor of this superoutburst but it was so widely separated  
from its main part by quiescence that the precursor looked 
like an isolated normal outburst, as discussed below. 
Thus this superoutburst has turned out to be an extreme case of 
precursor-main types in which the dip between the precursor and 
the main was so deep so as to touch quiescence. 

Let us now look at the enlarged light curve in figure 
\ref{fig:v1504spec2dfreqso8} (panel a).
We can see that a periodic hump most likely of ``superhump''
origin first appeared on the descending branch of the 
preceding normal outburst and it continued to quiescence.
To confirm that the signal was really of superhump origin,  
we made the 2D power spectral analysis of the light curve. 
We note here that frequencies of three periodic signals in V1504 Cyg are 
$\nu_{\rm orb}=14.38$c/d for the orbital frequency, 
$\nu_{\rm pSH}\sim 13.8$c/d, and $\nu_{\rm nSH}\sim 14.7$c/d, respectively. 
We show the power spectrum in figure \ref{fig:v1504spec2dfreqso8}
(panel c) and the frequency analysis with the 
phase dispersion minimization method (the PDM method \cite{PDM}) 
in figure \ref{fig:v1504spec2dfreqso8} (panel d). 
As seen in the figure, the signal of the positive superhump 
at a frequency around 13.5~c/d first appeared in the descending branch of 
the normal outburst at around 55876.5 and it continued till quiescence 
with a small amplitude.  This signal had a lower frequency than
the positive superhump during the later course of the superoutburst,
and corresponds to stage A (growing stage) superhumps in \citet{Pdot}.
A new outburst started at around the day 55879. 
The existing positive superhump grew in amplitude together with 
the increase of brightness and both the superhump and the system brightness 
 reached their maxima simultaneously at around the day 55879.5. 
 This demonstrates very clearly that the superoutburst was triggered by 
 the superhump (or by the tidal instability) as the enhanced 
 tidal dissipation in the eccentric (flexing) disk 
 rekindled the thermal instability 
 (from the cold state to hot one) at the outer edge of the disk and 
 a superoutburst was initiated by the tidal instability, giving a strong 
 support to the original thermal-tidal instability model (TTI model) 
 proposed by \citet{osa89suuma}. 
  
The same kind of phenomenon occurred in superoutburst No. 9 of V1504 Cyg. 
A preceding normal outburst occurred at the day 55991 in which the superhump 
with a frequency $\nu\sim 13.5$c/d first appeared in its descending branch 
and it continued to quiescence. 
The next superoutburst occurred 5~d after this 
normal outburst. Unfortunately, a strong noise (most likely caused by 
a solar flare) happened between 55994 and 55995, which deteriorated 
the Kepler data in the most delicate part of the very dip 
between the precursor normal outburst and the next superoutburst 
and it made further analysis impossible.

\begin{figure*}
  \begin{center}
%    \FigureFile(140mm,190mm){v344lyrspec2dfreqso.eps}
    \FigureFile(140mm,190mm){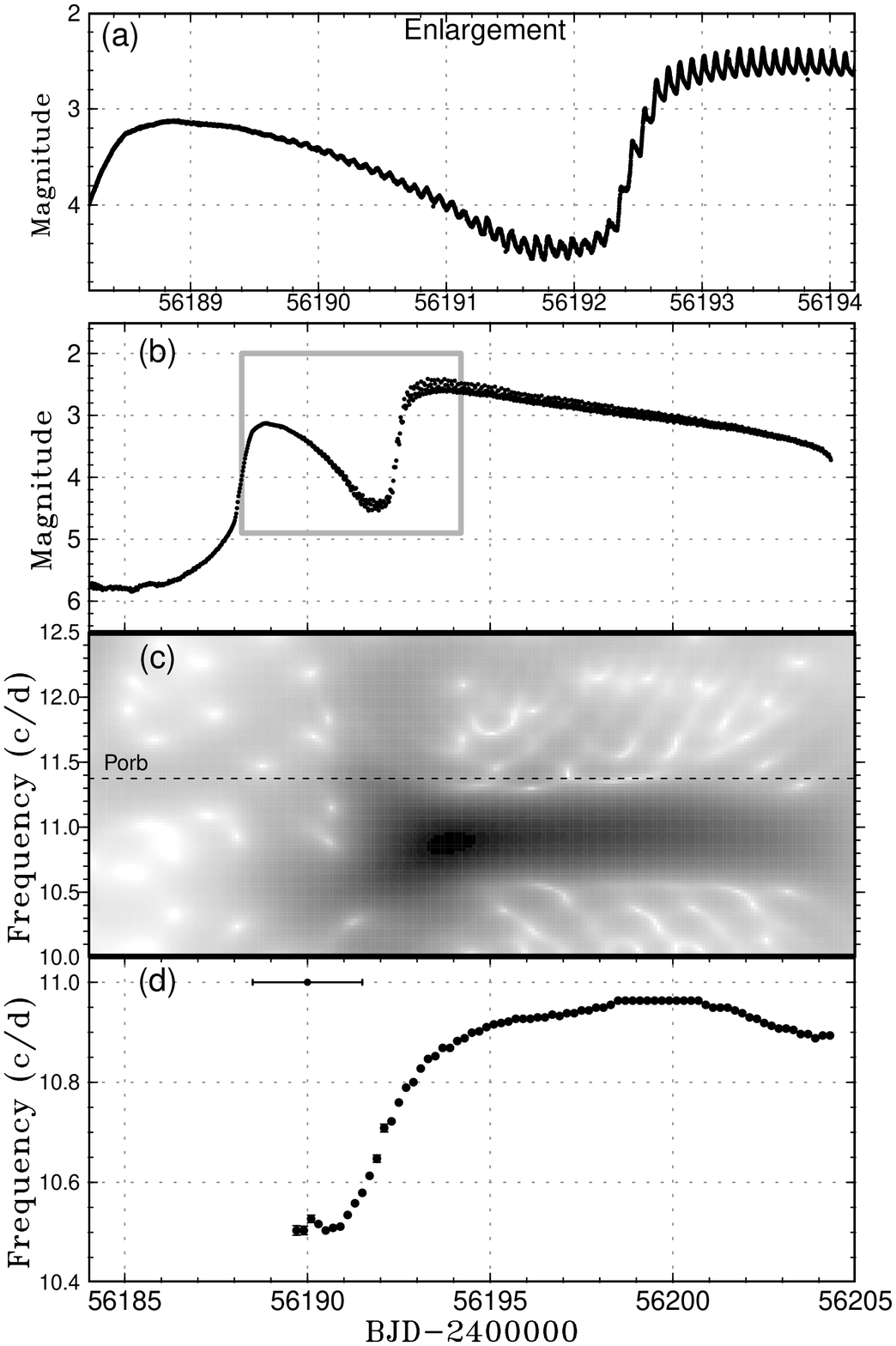}
  \end{center}
  \caption{
  Kepler light curve of V344 Lyr for superoutburst around BJD 2456190.
  (a) Enlargement of the box in panel b. The Kepler data were 
  binned to 0.001~d.  Superhumps developed
  between the short outburst and the superoutburst.
  (b) Light curve corresponding to the interval of panels b and c;
  the Kepler data were binned to 0.01~d
  (c) Two-dimensional power spectrum.
  The sliding window and the time step used are 5~d and 0.1~d,
  respectively.
  (d) Frequency variation of superhumps.
  The initial low frequency
  superhumps correspond to the growing stage superhumps (stage A
  superhumps).}
  \label{fig:v344lyrspec2dfreqso}
\end{figure*}

The same sort of phenomenon was also observed in V344 Lyr for 
a superoutburst which occurred at around 56193. 
In figure \ref{fig:v344lyrspec2dfreqso} we show the corresponding one 
for V344 Lyr.  As seen in this figure, the precursor normal outburst, 
which started around 56187, reached its maximum around 56188.5. 
The superhump with a frequency around 10.5~c/d appeared in 
the descending branch of this normal outburst. 
However, the system brightness did not reach quiescence 
but it reached a local minimum (the dip) and it began to grow together 
with the growth of superhump amplitude.  The dip between the precursor 
and the main outburst in this case was slightly shallower 
than that of V1504 Cyg, not reaching the quiescence. 

Let us now discuss a problem why some of precursor normal outbursts are 
so much separated from the main outburst -- by as much as 5~d. 
This can be understood in the following way.  The triggering normal outburst 
occurs rather randomly with respect to the 3:1 resonance 
radius (i.e., the tidal instability radius); sometimes the disk expands well 
beyond the 3:1 resonance radius ($r_{3:1}$) during 
the triggering normal outburst 
so that a large amount of disk mass is pushed into the instability region 
while sometime the disk expands just beyond the $r_{3:1}$ so that only a small 
amount of the disk mass is left within the instability region.  In the former 
case, the tidal eccentric instability is expected to be so strong 
that the superhump grows rapidly and it triggers the main superoutburst, 
producing a precursor-main superoutburst with a short interval 
between the precursor and the main outburst.  On the other hand, 
in the latter case even if the superhump grows in the descending branch of the 
normal outburst, mass addition with low specific angular momentum from the 
secondary star pushes the outer disk mass below the $r_{3:1}$ and it kills 
the tidal instability (and the superhump), resulting in the failed superhump 
discussed in the next subsection.  These are the two extreme cases 
and there must be an intermediate case in which the growth of the tidal 
instability is rather slow but it can survive during quiescence and it 
eventually triggers the main superoutburst, producing the precursor normal 
outburst well-separated from the main superoutburst discussed in this 
subsection.  We do not know how much days of separation between the precursor 
normal outburst and the main superoutburst are possible but observations 
indicate 4 to 5~d for high-mass-transfer ($\dot{M}$) systems like V1504 Cyg
and V344 Lyr.  In low-$\dot{M}$ systems, the interval between
the separated precursor outburst and the main superoutburst can be
even longer: 10~d in QZ Vir (1998, \cite{ohs11qzvir}) and
11~d in V699 Oph (2001, \cite{Pdot}).

In all superoutbursts in the Kepler light curves of V344 Lyr 
studied before in Paper II and by \citet{can10v344lyr}, 
the dip between the precursor and the main was so shallow that it was 
called ``shoulder'' by \citet{can10v344lyr}.  The latter authors 
tried to explain this phenomenon as a ``shoulder'' based on Cannizzo's 
pure thermal disk instability model. 
It may be clear that the superoutburst shown here with a deep dip 
can not be explained by \citet{can10v344lyr} 
model as already discussed in subsection 3.2 of Paper I.  

\subsection{Failed Superhump in a Normal Outburst Prior to a Superoutburst} 
\label{sec:failedSH}

\begin{figure}
  \begin{center}
%    \FigureFile(88mm,110mm){v1504failedsh.eps}
    \FigureFile(88mm,110mm){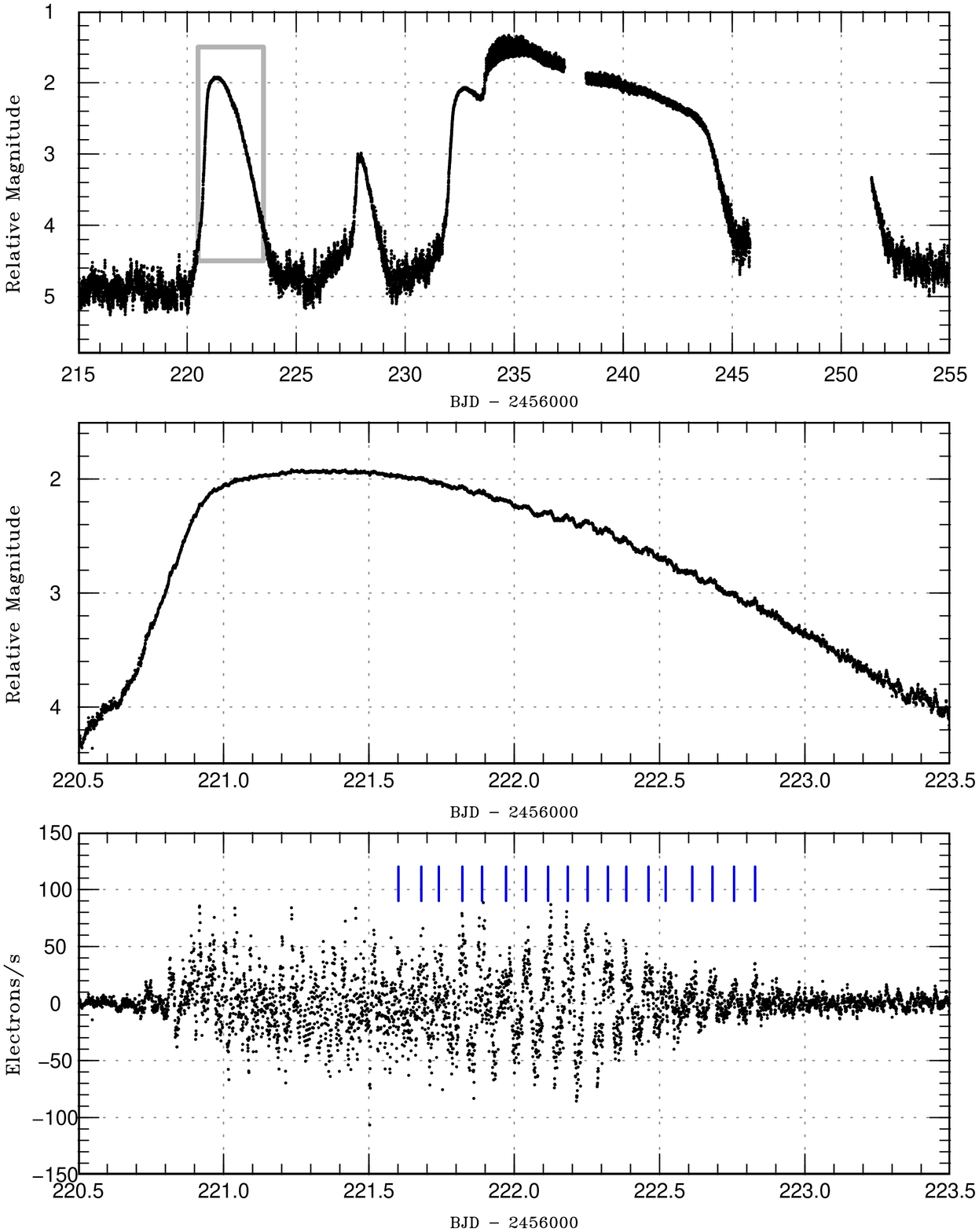}
  \end{center}
  \caption{Failed superhumps in V1504 Cyg.
  Upper: Light curve; the Kepler data were binned to 0.02~d,
  Middle: Enlarged light curve.  Transient superhumps were
  present on the descending branch.
  Lower: Residual signals in electrons s$^{-1}$.  The ticks
  represents the time of maxima of failed superhumps.}
  \label{fig:v1504failedsh}
\end{figure}

Another interesting phenomenon in the light curve of V1504 Cyg was 
a failed superhump seen in the descending branch of a normal outburst 
prior to a superoutburst.  Figure \ref{fig:v1504failedsh} shows such 
an example for a normal outburst which occurred around 56221. 
We can see very clearly that a signal of periodic light hump with 
a period around 14~c/d first appeared in the descending branch of 
this outburst but it disappeared before the system brightness 
reached quiescence. 
This signal was of the superhump nature because its frequency corresponds 
to that of the ordinary superhump and we call this the ``failed superhump''  
because it failed to excite a superoutburst.  This phenomenon is exactly the 
same as that of ``aborted superhump'' discussed by \citet{osa03DNoutburst}. 
The next superoutburst occurred at around 56233, 10~d after 
this normal outburst. 
It is very interesting to compare two light curves of 
figure \ref{fig:v1504spec2dfreqso8} and 
figure \ref{fig:v1504failedsh} where the former one is for 
the precursor normal outburst and the latter is 
for a normal outburst with the failed superhump. 
The superhump signal appeared in the descending branch of a 
normal outburst in both figures.  It continued until quiescence 
and it led to the superoutburst in the former case while 
it died down before quiescence in the latter case. 

\begin{figure}
  \begin{center}
%    \FigureFile(88mm,110mm){v344failedsh.eps}
    \FigureFile(88mm,110mm){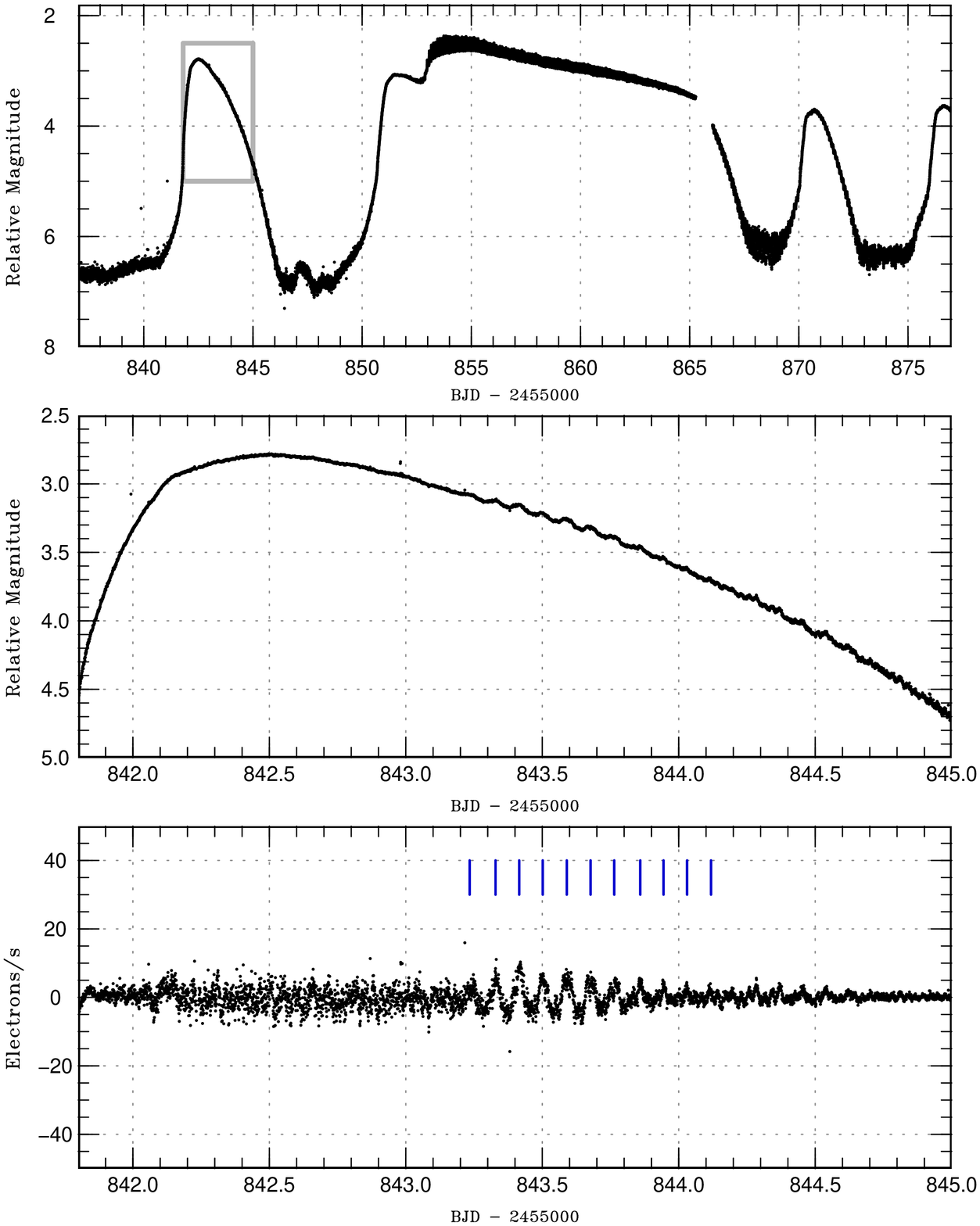}
  \end{center}
  \caption{Failed superhumps in V344 Lyr.
  Upper: Light curve; the Kepler data were binned to 0.02~d,
  Middle: Enlarged light curve.  Transient superhumps were
  present on the descending branch.
  Lower: Residual signals in electrons s$^{-1}$.  The ticks
  represents the time of maxima of failed superhumps.}
  \label{fig:v344failedsh}
\end{figure}

Figure \ref{fig:v344failedsh} shows another example of this phenomenon 
for V344 Lyr in a normal outburst at around 55842, which preceded 
the next superoutburst No. 8 by 8~d.  We also note that the same 
phenomenon was already noticed by \citet{Pdot3} 
for a normal outburst at around 50068 just prior to superoutburst 
No. 1 of V1504 Cyg in the Kepler light curve. 

The phenomenon of the failed superhump supports a picture presented 
by \citet{osa03DNoutburst} and thus supports the TTI model. 

\subsection{Mini-outbursts in the Type S Supercycle} \label{sec:miniNO}

One of unsolved problems in the Kepler light curves of V1504 Cyg and V344 Lyr 
concerns about the outburst intervals during a supercycle in the Type S 
supercycle.  As already pointed out by \citet{can12v344lyr}, the trend of 
quiescence intervals in the Kepler light curves of V1504 Cyg and V344 Lyr 
is to increase to a local maximum about half way through the 
supercycle and then to decrease back to a small value by the time of the 
next superoutburst.  The same trend was also noticed for the Type S 
supercycle of VW Hyi by \citet{sma85vwhyi}.
\citet{can12v344lyr} have argued that this is 
inconsistent with Osaki's thermal-tidal instability model.  We admit that 
this is one of difficulties in the TTI model 
and we examine this problem in this subsection. 
However, we note here that although \citet{can12v344lyr} criticized the TTI 
model, their own pure thermal instability model \citep{can10v344lyr}  
fails to reproduce this phenomenon as well and the two models 
are equal in this respect. 

Let us look in figures \ref{fig:v1504spec2d} and \ref{fig:v1504spec2dlasso}
at the light curve of supercycle No. 8 
of V1504 Cyg which is a typical Type S supercycle and in which 14 normal 
outbursts are recorded. 
We find that the quiescence intervals in the later half 
of the supercycle (from BJD 2455835 to 2455880) are definitely shorter 
than those in the middle as pointed out by \citet{can12v344lyr}. 
Furthermore the outburst amplitudes tend to decrease with advance of 
the supercycle phase and two mini-outbursts occurred during this period. 
The occurrence of frequent outbursts with low amplitudes in the later half of 
a supercycle suggests that the disk seems to know somehow its 
approach to the next superoutburst. 
A similar tendency is also seen in supercycles No. 9, 11, and 12, 
and No. 6 in the later half of which the signal of the negative superhump 
disappeared as shown in figure 7 of Paper II.  

By criticizing the TTI model, \citet{can12v344lyr} have mentioned 
that the observed trend of quiescence intervals indicates 
in the TTI model that the triggering radius for normal outburst (NO) 
does not move uniformly outward with each successive NO, 
but rather attains a local maximum and then recedes. 
This was clearly their misinterpretation of the TTI model. 
The TTI model predicts that the disk radius increases uniformly 
with each successive NO even in the later half of a supercycle.
The origin of decreasing interval between normal outbursts 
in the later half of the Type S supercycle must be sought 
in some other causes.  Observations mentioned above 
indicate rather that the disk's outer edge in the later half of 
a supercycle is approaching either to the 3:1 
resonance radius or to the tidal truncation radius.  Here we propose 
a possible scenario for occurrence of frequent normal outbursts with low 
outburst amplitudes in the later half of the Type S supercycle 
within the TTI model, although it is still very much speculative.  

We seek its cause in the radius dependence of the S-shaped thermal 
equilibrium curve in the thermal instability model.
In the thermal-viscous instability model there are two critical 
surface densities: 
$\Sigma_{\rm max}$  above which no cold state exists, 
and $\Sigma_{\rm min}$ below which no hot state exists. 
They are given by \citet{can88outburst} and \citet{war95book}: 
\begin{eqnarray}
\Sigma_{\rm max}=11.4r_{10}^{1.05}M_1^{-0.35}\alpha_{\rm C}^{-0.86} 
{\rm g \,cm}^{-2}, \\
\Sigma_{\rm min}=8.25r_{10}^{1.05}M_1^{-0.35}\alpha_{\rm H}^{-0.8} {\rm g \,cm}^{-2}, 
\end{eqnarray}
 where $r_{10}$ is the radius in units of $10^{10}$cm, 
 $M_1$ is the mass of the primary white dwarf in units of the solar mass, 
 and $\alpha_{\rm C}$ and $\alpha_{\rm H}$ are the viscosity parameters 
 in the cold disk and the hot disk, respectively. 
 It is generally thought that the viscosity in a hot ionized plasma 
 is given by the magneto-rotational instability 
 (MRI, see \cite{bal98ADreview}) 
 and $\alpha_{\rm H}\sim 0.1$.  On the other hand, the viscosity in the cold 
 state is still poorly known but in order to reproduce the dwarf nova outburst 
 by the thermal-viscous instability model, we need that $\alpha_{\rm C}$ is 
 much smaller than $\alpha_{\rm H}$.  Although no definite mechanism 
 to produce the viscosity in the cold state is yet known, the only mechanism  
 so far suggested is that of the tidal dissipation
 \citep{men00ADviscosity}.
 
 If the tidal dissipation is the only source for $\alpha_{\rm C}$, 
 we expect that $\alpha_{\rm C}$ may be a function of 
 the binary mass ratio, $q$, and the radial distance, $r$.
 A possible mass-ratio dependence of 
 $\alpha_{\rm C}$ is used to argue for a low value of $\alpha_{\rm C}$ 
 in WZ Sge stars to explain their long recurrence time.  Since the tidal 
 torque is strongly increasing with the radial coordinate and it increases 
 rapidly as the disk edge approaches the tidal truncation radius, 
 we expect that 
 $\alpha_{\rm C}$ is an increasing function toward 
 the tidal truncation radius.  The size of the loop of the S-shaped 
 thermal equilibrium curve depends on the ratio of 
 $\Sigma_{\rm max}$ to $\Sigma_{\rm min}$; the larger this ratio, 
 the larger is the S-shaped loop. If $\alpha_{\rm C}$ and $\alpha_{\rm H}$ 
 are assumed to be constant and have a definite ratio, 
 the S-shaped curve is self-similar 
 with respect to the radial coordinate, $r$.  This is the standard 
 assumption most researchers have adopted. 
 
 On the other hand, if $\alpha_{\rm C}$ has 
 a radial dependence while $\alpha_{\rm H}$ remains constant with respect to 
 the radius, the S-shaped loop becomes smaller as we approach near to the 
 tidal truncation radius.  A similar effect (a reduction in the ratio 
 of $\Sigma_{\rm max}$ to $\Sigma_{\rm min}$) may be produced 
 in the vicinity of the mass addition region due to the  associated 
 energy input (see, figure 2 of \cite{lin85CValphadisk}). 
 As a matter of fact, \citet{ich93SHmasstransferburst} have made light curve 
 simulations of SU UMa stars by using an $r$-dependent viscosity parameter, 
 $\alpha_{\rm C}$. The main purpose of their simulations was to reproduce 
 the ``outside-in''-type normal outbursts and its $r$-dependence 
 is rather mild, i.e.,  $\alpha_{\rm C}\propto r^{0.3}$. 
 In our scenario, we require a steep increase in $\alpha_{\rm C}$, 
 when the radius approaches to the tidal truncation radius 
 as compared with that used by \citet{ich93SHmasstransferburst}. 

 As discussed in Paper I, the normal outbursts in V1504 Cyg are thought to 
 be of ``outside-in'' in which the heating transition first starts 
 from the outer edge of the disk, propagating inward. 
 We now come to the main point of our scenario. In the Type S supercycle 
 in V1504 Cyg, as the disk expands outward with a successive normal 
 outburst in a supercycle and approaches to the tidal truncation radius, 
 the loop of S-shaped curve at the disk edge becomes smaller. 
 This may produce much frequent outbursts with a smaller outburst 
 amplitude in the later phase of a supercycle.
 Thus it is essential that the disk radius increases with advance of 
 supercycle phase in this scenario.   

\begin{figure}
  \begin{center}
%   \FigureFile(80mm,100mm){limit_cycle.eps} 
   \FigureFile(80mm,100mm){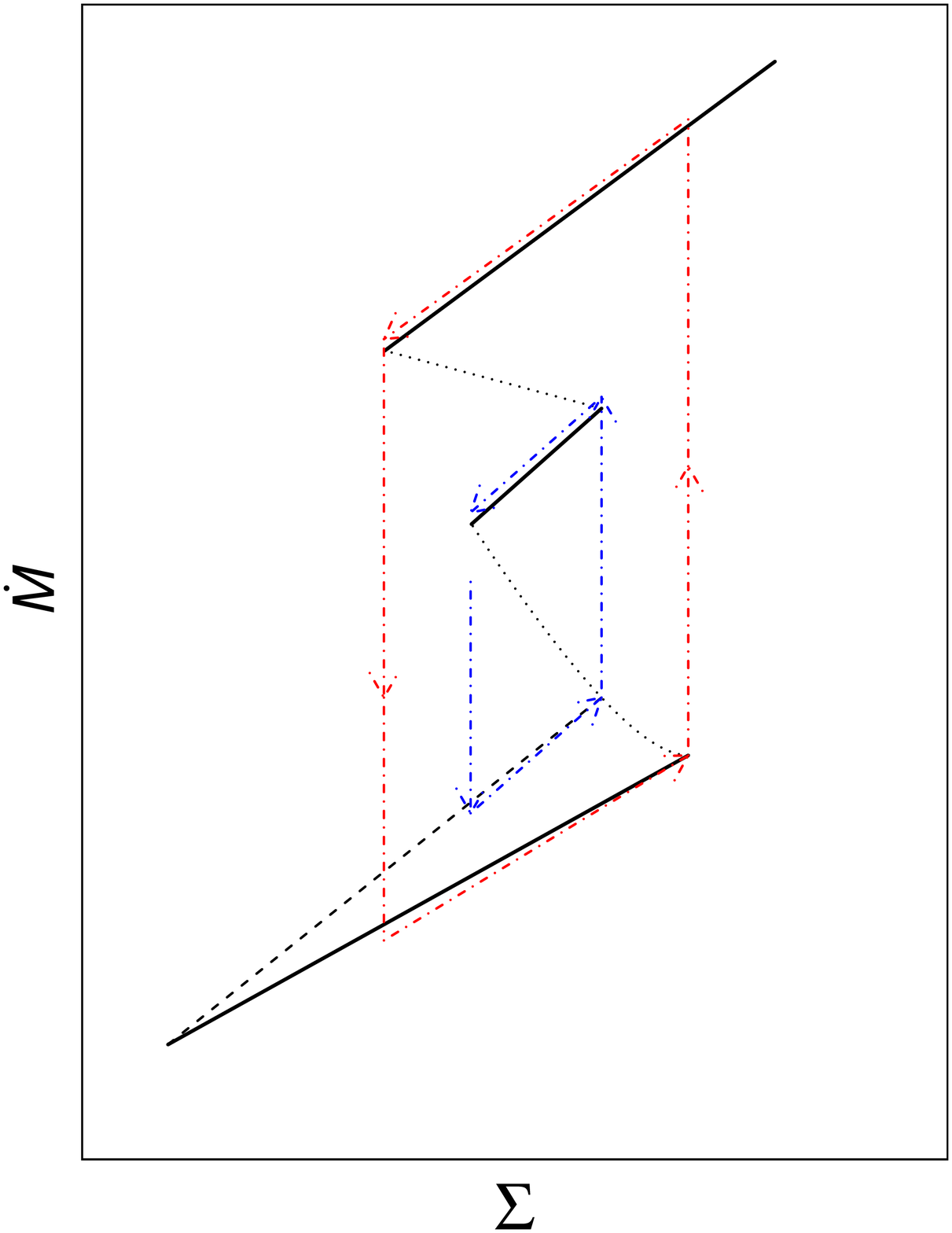}
   \end{center}
  \caption{The schematic picture of the thermal limit cycle instability with 
  a more complicated thermal equilibrium curve resembling the greek letter 
  $\xi$.  There exist three stable branches with positive slope; besides the 
  ordinary hot branch and cold branch an intermediate branch which is called 
  the warm branch. The ordinary thermal instability goes around the big loop 
  shown in the figure. Another limit cycle with a smaller loop 
  may occur if the cold branch is somehow elevated by an extra heating 
  possibily by the tidal dissipation.}
  \label{fig:limit_cycle}
\end{figure}

Let us now discuss about mini-outbursts which occurred in the later half of 
Type S supercycles of V1504 Cyg, that is, two mini-outbursts 
in supercycle (SC) No. 8, three in SC No. 9, two in SC No. 11, 
and one in SC No.6 which was not a typical Type S supercycle 
but in which the negative superhumps disappeared in its later stage. 
Here we propose a possible model which may explain  
mini-outbursts along this scenario.  In the standard disk-instability model 
we assume the so-called S-shaped thermal equilibrium curve for the thermal 
limit cycle instability.  However, a much complicated thermal equilibrium 
curve was obtained in the 1980s by various authors 
(particularly, \cite{min85DNDI}, \cite{mey87thermal}, \cite{can84ADvertical} 
among others) 
 and it looked more like the greek letter $\xi$ rather than the Roman 
letter S as shown schematically in figure \ref{fig:limit_cycle}
(see figure 4b of \cite{osa96review}).  In this type of thermal equilibrium 
curve, there are three stable branches; besides the ordinary 
hot and cold branches there exists an intermediate stable branch 
called the ``warm branch''.  In fact, \citet{min88uvdelay} has demonstrated 
that the stagnation stage due to this warm state can explain 
the UV delay observed in dwarf nova outburst. 

Let us explain our model for the mini-outburst by using the thermal 
equilibrium curve shown in figure \ref{fig:limit_cycle}.  We understand 
that the ordinary normal outburst is explained by the thermal limit-cycle 
instability in which the disk goes around the large loop shown in this figure. 
However, if the disk's outer edge approaches to the tidal truncation radius, 
an increased tidal dissipation may modify the cold branch of the equilibrium 
curve as shown by the dashed line of figure \ref{fig:limit_cycle}. 
It will then give rise to a premature ignition of thermal instability. 
In such a situation the disk may not jump to the hot state 
but it may be stopped at the intermediate warm branch, 
The resultant outburst may be of a small scale because the disk goes around 
the small loop (the small thermal-limit cycle) 
as shown in figure \ref{fig:limit_cycle}.  This is our explanation for 
mini-outbursts observed in V1504 Cyg.
In fact, the Kepler light curve of V1504 Cyg shows that the quiescence level  
was raised just prior to mini-outbursts, indicating some additional 
heating in quiescent disk.  As discussed above, 
besides the tidal dissipation, heating by collision of gas stream 
with the disk edge may also contribute to 
an additional heating in the type S supercycle.  

Observations of V1504 Cyg showed that mini-outbursts occurred rather randomly. 
This suggests that the extra heating in quiescent disk discussed above may 
occur rather radomly.  This suggests that the disk sometime goes around 
the large loop and the other time goes around the small loop.  However, we 
do not know any mechanisms resposible for such random behavior. 
Furthermore the triggering normal outburst was found to have a rather 
large amplitude.  These two points remain unsolved by this model. 
We leave them as a future problem to be solved. 

\section{Reply to Smak's Criticism to Our Paper} \label{sec:replytoSmak}

Recently, \citet{sma13negSH} published a paper in Acta Astronomica 
in which he challenged our main conclusions of our Paper I. 
He concluded in his paper that our main conclusions of 
Paper I were incorrect. 
Since his challenge is clearly serious for us, we present 
our detailed accounts to all of his criticisms in this section, 
by examining his criticisms one after another. 

\subsection{Amplitudes of Negative Superhumps}\label{sec:negAM}

In his subsection 2.2, \citet{sma13negSH} has argued 
that our results on the disk radius variation were inconsistent 
with observed amplitudes of negative superhumps. 
We think that {\it amplitudes} of negative superhumps 
have nothing to do with the disk radius variation as discussed below. 

\citet{sma13negSH} has tried to compare the amplitude of negative 
superhump during a superoutburst with that of quiescence. 
To do so, he assumed the light source of the negative superhump was 
solely due to the gas stream in a tilted disk both in quiescence 
and in a superoutburst. 
However we do not agree with him about the origin of 
light source of negative superhumps. 
We rather think that the light source of 
the negative superhump is most likely different between 
superoutburst and quiescence as evidenced from different wave forms 
between these two phases as shown in figures 7 and 8 in our Paper I. 
During superoutbursts the disk component will contribute to the light 
variation of negative superhump besides the gas stream component 
and we have discussed a possible origin of the wave form of the negative 
superhumps during a superoutburst in the Appendix of our Paper I. 
In fact, we showed in figure 5 of Paper II the variation in amplitude 
(in flux unit) of negative superhump together with the light curve 
and its frequency variation in a complete supercycle No. 5 of V1504 Cyg, 
which we studied in Paper I. 
As seen in the figure, the amplitude of negative superhump 
exhibited a characteristic variation.  Generally speaking, 
its amplitude (in flux unit) increases when an outburst occurs, 
which we interpreted as an evidence of contribution of 
the disk component to the negative superhump light source. 

Furthermore even if we accept Smak's assumption  
that the light source of the negative superhump were solely 
due to the gas stream in a tilted disk both in quiescence and 
in superoutburst, we reach an opposite conclusion to \citet{sma13negSH}
about the negative superhump amplitudes as discussed below. 
\citet{sma13negSH} has argued that the negative superhump 
amplitude must be smaller when the disk is 
larger and {\it vice versa}. However, the nSH amplitude must be 
determined by {\it difference} in the depth of potential well 
between the deepest arrival point of the gas stream 
and the shallowest point in a tilted disk.
What \citet{sma13negSH} referred to was the depth of 
potential well at the outer disk edge 
which corresponds to the shallowest point in the above discussion and 
he did not referred to the deepest arrival point. 
If the deepest arrival point is assumed to be 
the same in quiescence and in superoutburst in a tilted disk, 
the nSH amplitude must be larger when the disk's outer edge is larger, 
a conclusion exactly opposite to his. 

We think that Smak's criticism on amplitudes of negative superhump is 
irrelevant because light source of the negative superhump is most likely 
different between quiescence and a superoutburst and furthermore 
because his argument on the depth of potential well was incorrect. 

\subsection{Frequency Variations of Negative Superhumps during Normal 
Outburst Cycle} 

In his subsection 2.3, \citet{sma13negSH} criticized our results on 
frequency variations of negative superhumps during normal outburst 
cycles shown in our figure 5 in paper I. 
He stated that in our figure 5 the minima of $\nu_{\rm nSH}$ occurred 
$\sim 3$~d {\it before} the initial rise to outburst and the following 
increase of $\nu_{\rm nSH}$ till its maximum lasted for $\sim 3$~d 
while the model calculations show that expansion of the disk occurs 
nearly {\it simultaneously} with the rising light and lasts 
for only $\sim 0.5$~d (which corresponds to a viscous time scale).  

Since this was the most important criticism, we have already presented 
our detailed account on this criticism in subsection 2.2 of our Paper II. 
We do not repeat these discussions here but rather we would like 
to ask the readers to consult on our Paper II. 
Here we summarize the main points of our discussion of Paper II below.  
\begin{enumerate}
\item We made a mistake of 2~d in the time axis in figure 5 of 
the first version of astro-ph, arXir:1212.1516v1 (a preprint of Paper I) 
submitted on 2012 December 6, and this error was corrected 
in the second version submitted on 2013 January 6, and in the published 
version of PASJ.  We realized this error by Smak's criticism and 
we thank Dr. Smak for his careful scrutiny of our Paper I in 
its first version. 
\item The reason why the frequency jump in $\nu_{\rm nSH}$, 
in figure 5 of our Paper I, took 4~d [\citet{sma13negSH} said $\sim 3$~d]
instead of a much shorter 
time scale of $\sim 0.5$~d is due to a simple artifact 
in our calculations of local frequency with the window width of 4~d. 
\item By taking into account these two points, results shown in figure 5 
of our Paper I is consistent with the model calculations quoted 
by \citet{sma13negSH}. 
\end{enumerate}

\subsection{Negative Superhumps during Superoutbursts}

In his subsection 2.4, \citet{sma13negSH} criticized our results 
on the decrease in frequency of negative superhumps, $\nu_{\rm nSH}$, 
during the main part of 
superoutbursts indicating the decrease in the disk radius during the 
superoutbursts. 
On the other hand, \citet{sma13negSH} stated that 
the disk radius of Z Cha determined by himself from eclipses during 
superoutbursts remains constant throughout superoutburst. 

We believe that this is one of the most crucial issues to be tested   
in observations.  That is, the TTI model 
predicts the decrease in the disk radius during superoutburst 
while Smak's enhanced mass-transfer (EMT) model expects a more or less 
constant disk radius during superoutburst.

We note, however, some shortcomings in \citet{sma13negSH}.
These observations of Z Cha were performed only 1.5--6.5~d
after the onset of (different) superoutbursts, and did not cover
the full duration (10--12~d) of the plateau phase of superoutbursts
in this object.  The resultant error in his estimate of
the decrease in the disk radius was large [equation (2) in
\citet{sma13negSH}] due to the short baseline, and the decrease
in the disk radius in our Paper I data was only 1.6$\sigma$
different from his estimate.  It means that his data were not
statistically significant enough to test our results.
Furthermore, \citet{sma13negSH} used the times of ingress and
egress of the hot spot to estimate the disk radius.
Smak himself confessed that ingress and egress times do not
always give consistent results, and may not be directly used to
estimate the disk radius \citep{sma12zchaoycaroverflow}.

\begin{figure}
  \begin{center}
%    \FigureFile(88mm,80mm){iyumaradius.eps}
    \FigureFile(88mm,80mm){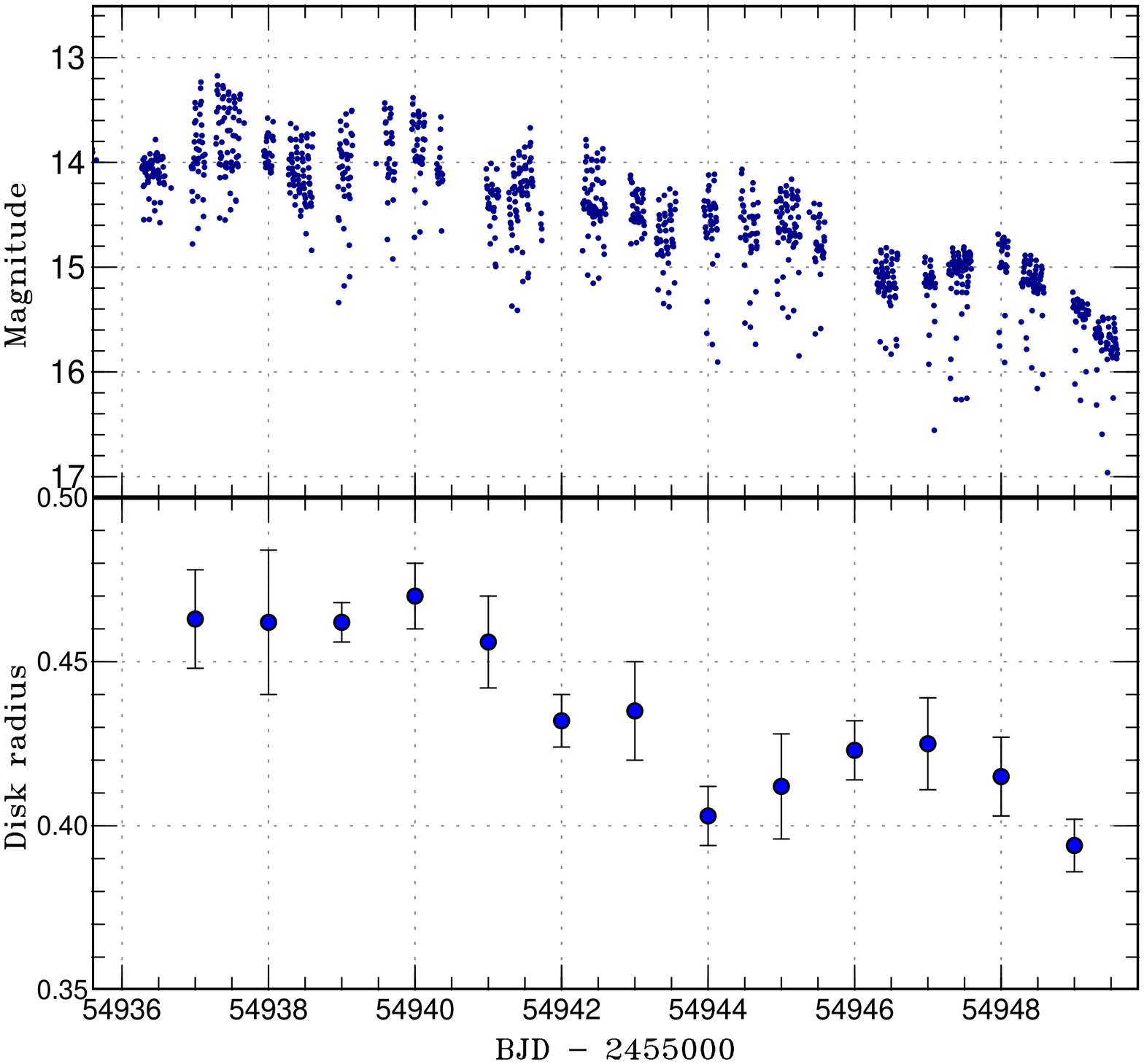}
  \end{center}
  \caption{Variation of disk radius during superoutburst
  in IY UMa (2009).
  (Upper): Magnitude binned to 0.005~d.
  (Lower): Disk radius obtained by modeling the eclipse profile.}
  \label{fig:iyumaradius}
\end{figure}

Eclipsing SU UMa-type dwarf novae, however, are a vital tool
in probing the variation of the disk radius in outburst,
and we show some clues from our own observations.
We used the 2009 observation of the eclipsing SU UMa-type dwarf nova
IY UMa (\cite{Pdot2}; \cite{Pdot3}) and estimated the disk radius by 
profile fitting.  We assumed that the disk is axisymmetric and ignored
the thickness.  We modeled the surface brightness of the disk
as a form of $\propto r^{-n}$, and obtained the model light curve.
By using Markov-chain Monte Carlo (MCMC) method,
we determined the disk radius and $n$ value.
In order to reduce the effect of the strong beat phenomenon between
the superhump and orbital periods, we averaged profiles for 4~d, 
where 2~d is the beat period.
We used $q$=0.125 and an inclination of 
86$^\circ$ \citep{ste03iyumaSTJ}.  The result is shown in figure 
\ref{fig:iyumaradius}.  Although our treatment is rough and based on
an assumption that the mean shape of the disk is axisymmetric,
the present result of eclipse analysis supports the decrease
in the disk radius during superoutburst.  We note that the
total duration of the eclipse is basically determined by
the radius of (the luminous part of) the disk, and the uncertainty
arising from the simplified model is relatively small.
This result may suggest that the disk radius did not decrease
strongly during the initial stage of the superoutburst,
when \citet{sma13negSH} observed Z Cha.
This issue has to be settled in observations in future. 

As stated clearly by \citet{sma13negSH}, the frequency variation of 
negative superhumps during the plateau stage of superoutbursts in 
V1504 Cyg, shown in figure 5 of our Paper I, indicates the decrease 
in the disk radius.  Since the disk radius increases secularly with 
a succession of normal outbursts from the end of the preceding 
superoutburst till the start of the next superoutburst as demonstrated 
in figure 5 of our Paper I, the disk radius must decrease during 
superoutburst from the standpoint of continuity argument and 
the observed variations in the nSH frequency mentioned above is 
consistent with this prediction. 

We have two more pieces of evidence for the decrease in disk-radius 
during superoutburst other than the variation in nSH frequency. 

The first evidence concerns the frequency 
variations of the {\it Positive Superhumps} during superoutbursts. 
In subsection 2.6 of our paper II, we studied the frequency variations of 
the positive superhumps in V1504 Cyg and V344 Lyr during superoutbursts. 
As shown in figures 9--12 of our Paper II, the positive superhump periods, 
$P_{\rm pSH}$,  (or the apsidal precession rates of eccentric disks) 
in V1504 Cyg and V344 Lyr show a rapid decrease from the highest 
$P_{\rm pSH}$ (or the highest apsidal precession rates) 
at the start of a superoutburst, to a less rapid decrease 
during the plateau stage of the superoutburst. In Paper II we interpreted 
the initial rapid decrease is due to propagation of eccentricity wave 
from the 3:1 resonance region to the inner region of the disk. 
The most interesting point in these Kepler observations  both in V1504 Cyg 
and V344 Lyr is a slower decrease in the positive superhump period 
during the plateau stage of the superoutburst. As discussed in subsection 2.6 
of Paper II, this is understood to be due to a decrease in the disk radius 
during the superoutburst. 

The second evidence concerns about luminosity level between the 
precursor maximum and the end of plateau stage of superoutburst 
when the rapid decline from superoutburst begins. 
The Kepler light curves of V1504 Cyg and V344 Lyr show 
that the former (the precursor maximum) is brighter typically 
by about 0.5 mag than the latter (the end of the plateau stage). 
According to the disk instability theory, these two stages are exactly when 
the cooling transition just starts at the outer edge of the disk. The critical 
effective temperature of the disk's outer edge when the cooling transition 
starts is well specified in the thermal instability 
theory to be around 7500K. Since the surface brightness in these two stages 
are more or less the same, difference in brightness by about 0.5 mag   
between these two stages is explained only by difference in the surface area 
of the disk. This indicates that the disk radius at the end of superoutburst 
is about 0.8 of that at the start of superoutburst. We thus reach a 
conclusion that the disk must contract during superoutburst. 
This is a very simple argument free from any sophisticated theory and  
we can easily confirm this just by looking at the Kepler light curves 
of these stars by eye. 

Here we must add the potential caveat of this argument. 
In the above discussion, we assumed implicitly a steady hot disk both 
at the precursor maximum and at the end of the plateau stage. 
The assumption of the steady disk is a good approximation for the end of 
the plateau stage but it is not for the precursor maximum. At the 
precursor maximum, the disk is non-steady as the heating front may still 
be propagating inward in the disk.  This affects the above discussion 
in two different ways; firstly the part of the disk may be still in cold state 
(i.e., the disk is not fully hot) and the hot part may not be a circular 
disk but rather a circular ring. The second effect concerns the existence of 
heating front accompanied with narrow spikes in the surface density 
and the temperture distribution within the disk may differ 
from that of the steady disk.  

\subsection{Comparison with Other Systems} 

In his subsection 2.5, \citet{sma13negSH} examined negative superhumps 
and their variable periods in several other 
dwarf novae in the literature. He concluded that (1) the decreasing 
nSH period during supercycles is a common phenomenon among dwarf novae 
with superoutbursts, while (2) the nSH period variations during their normal 
outburst cycles  occur only in some of them. 
In particular, nSH period variations observed in V1504 Cyg cannot 
be considered as representative for all such systems. 

Here we study these two points raised by Smak, by examining 
individual systems in more detail. The systems quoted by him were 
V503 Cyg, BK Lyn, V344 Lyr, and ER UMa. 
It has turned out all five systems showing negative superhumps  
are those SU UMa stars with rather short supercycles. Since  
the behaviors of the negative-SH period variations in these systems 
have a strong correlation with the supercycle properties, we 
first summarize their properties in table \ref{tab:interval}. 
The first column (1) of the table is star's name, 
(2) $T_S$: the supercycle length, (3) $T_S^*$: the supercycle length 
excluding the superoutburst, that is, the interval from the end 
of a superoutburst to the next superoutburst, 
(4) $T_N$: the normal-outburst interval or the cycle length 
of normal outbursts, and (5) $T_Q$: 
the quiescence interval. 
All quantities listed in table \ref{tab:interval} refer 
to the Type L supercycle. We make this remark 
because the normal-outburst intervals are 
very different between the Type L supercycle and the Type S supercycle as 
discussed in Paper I. 

\begin{table}
\caption{Various intervals in units of days for SU UMa stars showing 
negative superhumps with the Type L supercycle.}
\label{tab:interval}
\begin{center}
\begin{tabular}{p{60pt}p{35pt}p{35pt}p{30pt}p{30pt}}
\hline
name & $T_S$\commenta & $T_S^*$\commentb & $T_N$\commentc 
& $T_Q$\commentd \\
\hline
V1504 Cyg & 110 & 98 & 16 & 12 \\
V344 Lyr & 102 & 85 & 10 & 6 \\
V503 Cyg & 89 & 70 & 30 & 25 \\
ER UMa  & 50 & 17 & 7 & 2 \\
BK Lyn  & 45 & 20 & 5 & 0 \\
\hline
\multicolumn{5}{l}
{\commenta   $T_S$: the supercycle length} \\
\multicolumn{5}{l}
{\commentb  $T_S^*$: the supercycle length excluding superoutburst} \\
\multicolumn{5}{l}
{\commentc  $T_N$: the interval of two consecutive normal outbursts}\\
\multicolumn{5}{l}
{\commentd  $T_Q$:  the quiescence interval} \\
\end{tabular}
\end{center}
\end{table} 

Let us now examine nSH frequency variations for individual stars by 
taking into account their supercycle properties listed in 
table \ref{tab:interval}. To do so, we use more recent data 
than those of \citet{sma13negSH}, when they are available, As discussed 
by \citet{sma13negSH}, we discuss two phenomena, separately: 
(1) secular variation of nSH frequency with supercycle phase,  
and (2) nSH frequency (or period) variations during normal outburst cycles.

As for the nSH frequency variations during normal outburst cycles, 
we expect that the frequency difference between an outburst and quiescence, 
$\delta \nu=\nu_{\rm max}-\nu_{\rm min}$, will be larger if the normal 
outburst interval, $T_N$, (or $T_Q$) is longer because mass added 
by gas stream having lower specific angular momentum is more accumulated 
in the disk during quiescence if the quiescence interval is longer. 
This can be confirmed in figure 5 of Paper I for supercycle No. 5 
of V1504 Cyg where five normal outbursts occurred. 
The longest quiescence interval occurred between the third and the fourth 
normal outbursts and the jump in the nSH frequency from minimum to maximum 
was the largest for the fourth normal outburst. If we examine table 
\ref{tab:interval}, we find that the cycle length of normal outbursts of 
ER UMa and BK Lyn are as short as 7~d and 5~d, respectively. 

The time variations in nSH frequency (or period) have been studied 
by two different methods: (1) variations of local frequency of nSH 
with time, and (2) study of the $O-C$ diagram for times of superhump maxima. 
The first method is more suitable for frequency variations 
on a short time-scale while the second method is more suitable for long 
time-scale frequency variations because the $O-C$-diagram represents 
an integral form of the frequency variation. We do not expect 
any large variation in the $O-C$ diagram during an outburst cycle 
as short as several days seen in ER UMa and BK Lyn. 

Let us examine individual stars more closely. 
We start from V344 Lyr. In our Paper II we have studied frequency 
variations of negative superhumps by using the Kepler data of V344 Lyr, and 
we showed our results in Figure 9 of our Paper II  for supercycle No. 7 
and in Figure 10 for supercycle No. 2 of V344 Lyr. 
 From these two figures we find that V344 Lyr exhibited basically 
the same pattern of frequency variation in the nSH  
as that of V1504 Cyg, that is, (1) the nSH frequency increases 
when an outburst occurs while it decreases during quiescence, and (2) the 
cycle averaged nSH frequency increases secularly with advance of supercycle 
phase. However, the amplitude of frequency variation within an outburst 
cycle in V344 Lyr is smaller than that of V1504 Cyg. This difference is most 
likely caused by difference in the quiescence intervals, $T_Q$ in table 
\ref{tab:interval} in these two stars. V1504 Cyg has a longer quiescence 
interval than V344 Lyr and it exhibited more clear-cut variation in nSH 
frequency than V344 Lyr. We also note that our results are consistent 
with those of \citet{woo11v344lyr}. 

As for V503 Cyg, this system has been in the Type S supercycles in recent 
years (\cite{Pdot4}; \cite{pav12v503cyg}; \cite{Pdot5})
and no new data for the negative superhumps in this 
system are unfortunately available. Only available data are from 
\citet{har95v503cyg}, and \citet{sma13negSH} summarized their results 
as follows: the negative SH frequency in V503 Cyg was higher in 
outburst and lower during quiescence but no secular variation with 
supercycle phase was seen. As seen table \ref{tab:interval}, V503 Cyg is 
unique because it has a shorter supercycle length than that of V1504 Cyg but 
it has a longer quiescence interval. Observations by \citet{har95v503cyg} 
for variation of nSH periods between outburst and quiescence 
are consistent with our expectation for the negative SH frequency variation 
because V503 Cyg had long quiescence intervals.  However, it might 
have been difficult to find its secular variation in \citet{har95v503cyg} 
because of a shorter supercycle length in V503 Cyg.  
We hope that this point may be clarified by future observations sometime 
when this star enters in the Type L supercycle.   

BK Lyn was a novalike variable with a small brightness variation of 
about 0.2 mag until early 2000
(see, \cite{ski93bklyn}; \cite{Pdot4}; \cite{pat13bklyn}). 
However, it has begun to show dwarf nova eruptions as its mass-transfer 
rate apparently decreased and it is now classified as 
an ER UMa star,\footnote{
   The object again returned to a novalike state in 2013 April
   \citep{Pdot5}.
}
a subclass of SU UMa stars with a short supercycle length less than 50~d 
and extremely short normal outburst cycle of around 5~d
(\cite{kat95eruma}; \cite{rob95eruma}).
BK Lyn showed clear supercycles of around 45~d. It exhibited both 
positive and negative superhumps.  The $O-C$ diagram for negative SH was shown 
in figure 26 of \citet{Pdot4} which exhibited a concave upward pattern, 
indicating that nSH period decreased 
(i.e., the frequency increased) between two successive superoutbursts, 
as already noted by \citet{sma13negSH}. This means that 
the disk radius increases secularly with advance of a supercycle phase 
between two superoutbursts while it decreases during a superoutburst. 
However, no clear evidence is seen in the same diagram about any decrease in 
frequency in normal outburst cycles.
This is most likely due to a short interval of two consecutive normal 
outbursts in BK Lyn as compared with that of V1504 Cyg. 

\begin{figure*}
  \begin{center}
%  \FigureFile(150mm,120mm){bklynnegper.eps}
  \FigureFile(150mm,120mm){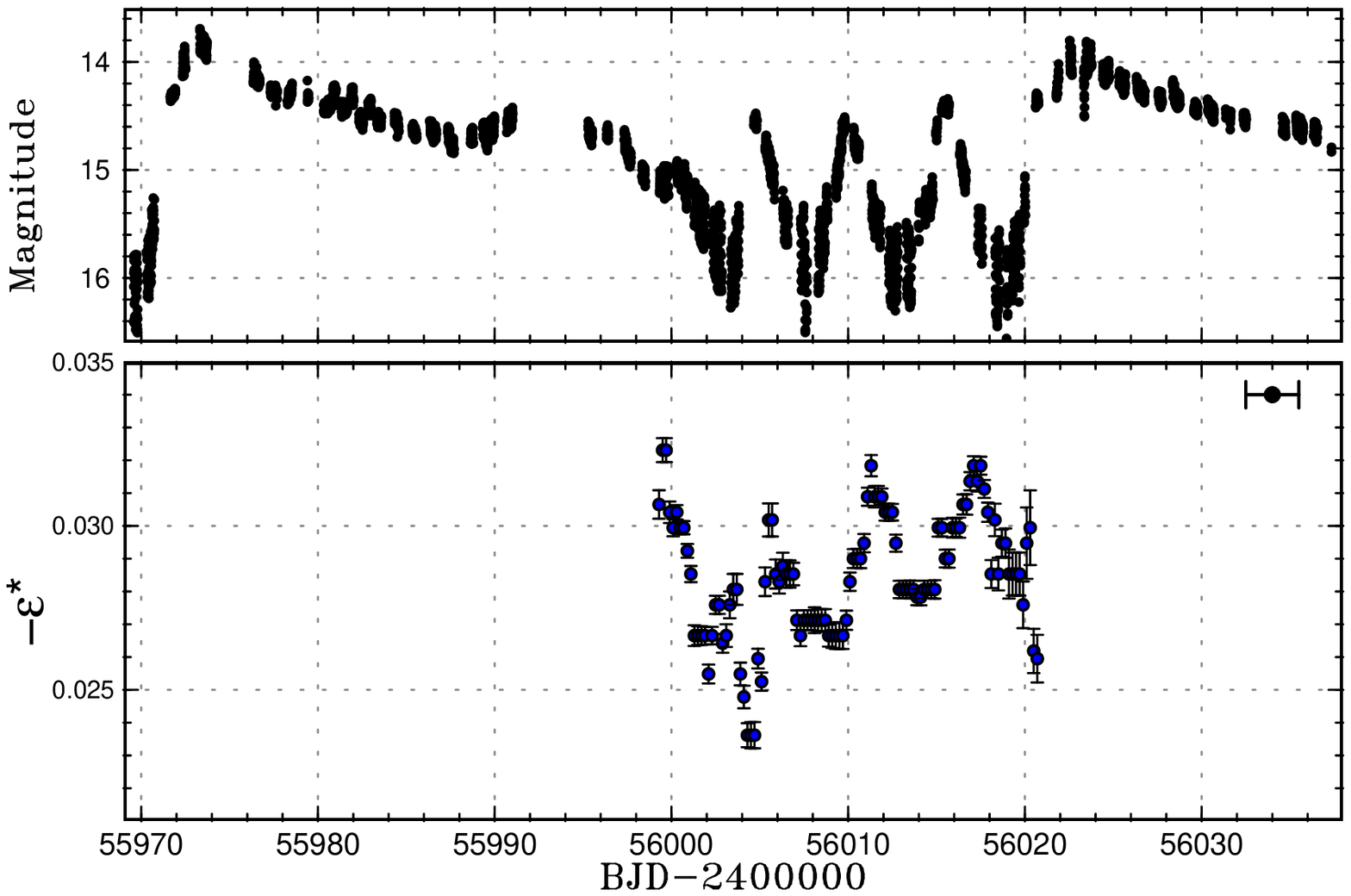}
     \end{center}
  \caption{Time evolution of frequency of the negative superhump 
  during an inter-superoutburst of BK Lyn.
  The upper panel shows light curve while the lower panel does variation 
  in the nSH frequency excess $\epsilon^*$ over the orbital frequency. 
  The frequency (or period) was calculated by using the PDM method 
  with a window width of 3~d and a time step of 0.2~d. The window width is 
  indicated as a horizontal bar at the upper right corner of the lower panel.}
  \label{fig:bklynnegper}
\end{figure*}

By using a combined set of the data used in \citet{Pdot4} and 
the additional data supplied by E. de Miguel (private communication 2013),  
we have examined local frequency variations of nSH in BK Lyn by using the 
PDM method for 
period determination.  Figure \ref{fig:bklynnegper}
illustrates our results in which a quantity $\epsilon^*$ representing 
the nodal precession rate over the orbital frequency is shown together 
with the light curve.  It clearly exhibits that the nSH frequency increases 
when an outburst occurs and it decreases when the star fades in brightness.
A Lasso two-dimensional spectrum of this star will be 
presented in \citet{Pdot5}.

ER UMa basically exhibits the same pattern as that of 
BK Lyn (\cite{ohs12eruma}; T. Ohshima et al. in preparation).
Thus we conclude that the disk radii in BK Lyn and ER UMa secularly increase 
with advance of supercycle phase from the end of a superoutburst to the 
next while they decrease during superoutbursts. In fact, 
\citet{dem12erumasass} demonstrated that the $O-C$ diagram 
for times of maxima of nSH waves 
of ER UMa showed a cyclic variation with the same cycle length 
as the supercycle itself. If this variation of nSH frequency is interpreted 
as the disk radius variation, 
it indicates that the disk radius decreases during superoutbursts  
and it increases secularly during inter-superoutburst, 
completely consistent with the 
prediction of the thermal-tidal instability model.  

For all SU UMa stars, V344 Lyr, BK Lyn, and ER UMa,  
observed variations in the nSH frequency 
exhibited basically the same pattern during supercycles: 
(1) the secular increase in nSH frequency between the two successive 
superoutbursts. (2) On the other hand, the decrease 
in nSH frequency during normal outburst cycles depends very much 
on the cycle lengths of normal outbursts in a sense 
that the longer the outburst interval, the larger the frequency decrease, 
In particular, two stars, ER UMa and BK Lyn, did not show any clear variation 
during outburst cycles in the $O-C$ diagrams 
because these two stars have extremely short outburst cycles. 
However our new analysis for variations of nSH frequency for BK Lyn 
by using the PDM method for the period 
determination showed clear up and down in nSH frequency during outburst 
cycles, which is consistent with that of V1504 Cyg. 

  We conclude that observed nSH frequency variations during 
supercycles in these 5 SU UMa stars are all consistent 
with the prediction of the thermal-tidal instability model. 
Quite recently, yet another example showing the same pattern
(KIC 7524178 = KIS J192254.92$+$430905.4) has been identified
\citep{kat13j1922}, strengthening this conclusion.

\subsection{Negative Superhumps and Their Variations}

In his subsection 2.6, \citet{sma13negSH} discussed the variation in 
negative superhump periods and criticized our Paper I by arguing that 
many different causes are responsible for the variation in negative 
superhump periods while \citet{osa13v1504cygKepler} limited 
their discussion only to the disk radius variation. 
Here we examine his criticism.

Concerning about many different causes, his discussions consisted 
of two different points.

\begin{enumerate} 
\item 
The first point concerns the nodal precession period, $P_{\rm prec}$,   
The retrograde precession period 
of a tilted disk depends not only on the disk radius but also 
on the distribution of its surface density $\Sigma(r)$ 
(cf., \cite{lar98XBprecession}, \cite{mon09diskprecession}).
Thus the variation in the surface density 
distribution has to be taken into account when variation of the negative 
superhump period is discussed. 
\item The second point concerns the observed nSH period. 
He argued that the observed nSH period is determined by the interval of time 
between two successive maxima resulting from the collision of the stream with 
the surface of the tilted disk and thus it depends also on the flight time of 
the stream elements from $L_1$ to the effective point of collision. 
He argued that the effects of the flight time have to be taken into account 
for the nSH period variation.
\end{enumerate}
We discuss below these two points separately.

\subsubsection{The effects of mass distribution in the disk on the nodal 
precession rate in a tilted disk} 

Concerning about the first point raised by \citet{sma13negSH}, 
we completely agree with him and here we discuss it.

We have already discussed 
the effects of different surface-density [$\Sigma(r)$] distribution 
on the nodal precession rate of a tilted disk in subsection 2.5 and Appendix 
of Paper II. There we introduced the precession rate of a tilted disk 
over the orbital frequency by $\epsilon^*_{-}=\nu_{\rm nPR}/\nu_{\rm orb}$, 
where $\nu_{\rm nPR}$ and $\nu_{\rm orb}$ are the nodal precession frequency   
and the orbital frequency, respectively. This quantity is expressed 
(see, \cite{lar98XBprecession}) by 
\begin{equation}
\epsilon^*_{-}=\frac{\nu_{\rm nPR}}{\nu_{\rm orb}}
=1-\frac{\nu_{\rm nSH}}{\nu_{\rm orb}}
=-\frac{3}{7} \frac{q}{\sqrt{1+q}} 
(\frac{R_d}{A})^{3/2} \cos \theta.  
\label{equ:noprecession}
\end{equation}
Here the negative sign of 
$\epsilon^*_{-}$ signifies retrograde precession.
Since this expression is derived under an assumption of 
a certain mass distribution in the disk, we introduced in Paper II 
a correction factor $\eta$ to this expression for allowing a different 
mass distribution; 
\begin{equation}
\frac{\nu_{\rm nPR}}{\nu_{\rm orb}}
=-\eta\frac{3}{7} \frac{q}{\sqrt{1+q}} 
(\frac{R_d}{A})^{3/2} \cos \theta. 
\label{equ:noprecession_mod}
\end{equation} 
The correction factor $\eta$ was calculated for several mass distributions  
in the Appendix of Paper II and it was listed in Table 3 in that paper. 

Let us now discuss the variation of nSH frequency in supercycle No. 5 of 
 V1504 Cyg shown in figure 5 of Paper I. Since this figure was shown both 
 in Paper I and Paper II, we do not here reproduce it any more. Instead we 
 summarize its variation in what follows:  the nSH frequency 
 jumps from local minimum to local maximum every time when a normal outburst 
 occurs and it decreases monotonically during quiescence between two 
 consecutive normal outbursts. Furthermore the peak frequency just after a 
 normal outburst increases monotonically with a succession of normal outbursts 
 from the end of the previous superoutburst until the next superoutburst. 
 
 The most important discovery of Paper I is the secular variation of 
 the peak frequency of nSH during a supercycle of V1504 Cyg. 
 As far as the peak frequency of the nSH is concerned, the correction factor 
 $\eta$ takes a same value and it is given by $\eta\simeq 1.22$ 
 because the surface density distribution just after 
 a normal outburst is self-similar and it is approximately 
 given by $\Sigma \propto r^{1}$. 
 Thus we can conclude that secular variation of the peak frequency of nSH 
 during a supercycle reflects the disk-radius variation because $\eta$ 
 remains constant and thus 
 the disk radius just after an end of normal outburst 
 secularly increases monotonously with advance of supercycle phase, 
 which is in good agreement with a prediction of the TTI model. 
 
 As for the variation in nSH frequency during two consecutive outbursts, 
 we admit some uncertainty in the surface density distribution and so 
 in $\eta$ because we do not know either how mass is supplied 
 to the different part of the disk in a tilted disk 
 nor how the viscous diffusion modifies the surface density distribution 
 in quiescence. Nevertheless we think that the observed variation 
 of nSH frequency may reflect more or 
 less the disk-radius variation even in quiescence. The reason why we 
 think so is two-fold. Firstly, if we interpret the frequency variation 
 of nSH observed in V1504 Cyg solely due to variation in disk radius, it is 
 reminiscent of the disk-radius variation in an outburst cycle of U Gem 
discovered by \citet{sma84ugemdiskradius}, one of the most important 
discoveries in the history of the disk instability theory. 
Secondly, the surface density distribution in 
quiescence has a certain limitation because it can not exceed the local 
critical surface density $\Sigma_{\rm max}(r) \propto r^{1}$ 
anywhere in a cold disk. Because of this constraint, 
it is difficult to consider a situation in which the correction 
factor $\eta$ is very different from that of the start of quiescence. 
We may therefore conclude that the variation in nSH frequency 
during a supercycle of 
V1504 Cyg reflects more or less variation in the disk radius even 
if the effects of variation in the surface-density distribution are taken 
into account. 

\subsubsection{The effect of flight time of gas stream} 

On the other hand, we believe 
that the second point raised by \citet{sma13negSH} is clearly 
a problem of different (higher) level of approximations from 
the first point because the second point goes deep into 
a particular mechanism for the origin 
of light source of the negative SH. 
As discussed in subsection \ref{sec:negAM}, 
we also consider other possibility for light source of the nSH 
different from the gas stream component. 
When we study variation in characteristic period (or frequency) of an 
astrophysical object, the first step will be to examine variation of 
the underlying clock, in our case the precession rate of the tilted disk. 
Going deep into the origin of light source will be the next higher step.
We believe that our discussion on variation in the nSH frequency is still 
in the first step. 

Furthermore, even if we accept Smak's view about the origin of negative 
superhump light source as due to gas stream, we think 
that the effect of the flight time discussed by \citet{sma13negSH} 
will be negligible in the frequency variation of the negative superhumps 
as shown below.

  If we assume that the light
maximum of the negative superhump coincides with the epoch
when the gas stream hits the innermost part of a tilted disk.
The maximum difference of the flight time of the gas stream
then will not exceed the travelling time of the infalling
gas across the disk radius.  Assuming a maximum disk radius
of 0.46 $A$, we can evaluate the time the trajectory
falling from the L1 point within this radius to be 
0.132 $P_{\rm orb}$ for $q=0.22$.
The time of 0.066 $P_{\rm orb}$ elapses when the trajectory
reaches the periastron since it first enters the radius
of 0.46 $A$.  We can use the same upper limit of 0.132 $P_{\rm orb}$
as the variation of the flight time.  This value (0.009~d
in V1504 Cyg) can be directly compared to the $O-C$ diagram of 
the negative superhumps (figure 5 in \cite{osa13v344lyrv1504cyg}) 
since this flight time effect corresponds to the global $O-C$ variation.
The real $O-C$ variation is much larger, even restricting to
the variation in relation to the normal outburst, indicating
that Smak's second effect cannot explain the period variation
of negative superhumps.

\subsection{Evidence for the Enhanced Mass-transfer Model}

In his section 3, \citet{sma13negSH} stated that observational 
evidence accumulated showing prominent hot spots during superoutbursts, 
in particular in the case of Z Cha \citep{sma08zcha}. 
However, observational evidence for the enhanced mass transfer 
quoted by \citet{sma08zcha} is 
very much dependent on an assumption used there, that is, an assumption 
that light source during superoutbursts 
consisted of the axi-symmetric disk component and of the hot spot. However, 
quite a different conclusion is obtained by starting from a different 
assumption.  

\citet{osa03DNoutburst} questioned the interpretation of evidence for 
enhanced hot spot and they argued that the so-called orbital hump 
(or ``hot spot'') during superoutbursts simply results 
from the non-axisymmetric tidal dissipation pattern in eccentric disk 
and the observed eclipses are not of mass-transfer hot spot but 
rather of the superhump light source itself. Based on the smoothed-particle 
hydrodynamics (SPH) simulations of tidally unstable disks, 
\citet{tru05masstransfer} showed that the brightening can be attributed to 
tidal heating in eccentric disk, with no need for an increase in mass-transfer 
rate.  We believe that accretion disks during superoutbursts are far from 
 axi-symmetry and the non-axisymmetric tidal dissipation pattern must 
 be taken into account even in superhump phases far away from the superhump 
 light maximum.

This indicates that by starting from a different assumption, one gets a 
quite different conclusion.  In such a situation a real problem will be   
which assumptions are more appropriate in such a case. 
We leave the judgment to readers and to future research. 

The Kepler light curves of V1504 Cyg and V344 Lyr have demonstrated 
that the rise to superoutburst maximum and the growth of superhump occurred 
almost simultaneously and both of them reached their maxima at the same time. 
The enhanced mass-transfer model seems to contradict with this observation 
because the enhanced mass transfer causes a contraction of the disk radius 
away from the 3:1 resonance by adding extra mass with low specific angular 
momentum, and it kills the tidal instability and the superhump 
as already discussed by \citet{ich93SHmasstransferburst}, 
\citet{lub94impact}, and \citet{tru05masstransfer}.

\section{Summary} \label{sec:summary}

(1) We made a supplemental study of the superoutbursts and superhumps 
of V1504 Cyg and V344 Lyr by using the recently released Kepler data. 
We have basically confirmed the results given in Paper I and Paper II. 
In addition to them, we have found that the supercycle lengths of the  
type L supercycles are shorter than those of the Type S supercycles 
in V1504 Cyg. 

(2) The superoutburst No. 8 of V1504 Cyg was preceded by a precursor normal 
outburst which was well separated from the main superoutburst. The ordinary 
superhump first appeared during the descending branch of the normal outburst 
and it continued to quiescence and it began to grow in amplitude 
with the growth of the system brightness and these two reached simultaneously 
their maxima (the main superoutburst maximum). 
This superoutburst is understood as 
an extreme case of the precursor-main type in which the precursor and the 
main were separated by the quiescence. This demonstrates that the 
superoutburst is triggered by the superhump (i.e., the tidal instability), 
supporting the thermal-tidal instability model. 
A similar phenomenon was observed in V344 Lyr.

(3) In one of the normal outbursts just prior to the next superoutburst of 
V1504 Cyg, a signal of ``ordinary superhump'' nature appeared in its 
descending branch but it disappeared before quiescence and thus failed 
to trigger a superoutburst. 
This phenomenon called the failed superhump is exactly the 
same as that of ``aborted superhump'' discussed by \citet{osa03DNoutburst}.  
In V344 Lyr, on the other hand, superoutbursts were found to be preceded 
always by one to three normal outbursts in which impulsive negative superhumps 
appeared during the descending branch but they disappeared before quiescence. 

(4) We discussed mini-outbursts which occurred exclusively in the later half 
of the type S supercycles of V1504 Cyg. 
We proposed a possible scenario wherein  
the outburst intervals and the outburst amplitudes decrease in the later half 
of the Type S supercycles in V1504 Cyg. 

(5) \citet{sma13negSH} criticized our paper I \citep{osa13v1504cygKepler} 
in which he argued that our conclusions of Paper I were incorrect. 
We presented our replies to almost of all his criticisms by offering clear 
explanations to his criticisms. In particular, we presented more evidence for 
the frequency variation in the negative superhump in other SU UMa stars. 

(6) The study of Kepler light curves of two SU UMa stars, V1504 Cyg and 
V344 Lyr presented in this paper and our Papers I \& II 
demonstrates that almost all of the observational evidence supports 
the TTI model but it seems against the EMT model. We conclude that 
the TTI model is the only viable model for the superoutbursts and 
superhumps in SU UMa stars, in particular in V1504 Cyg and V344 Lyr. 

\medskip 
Note added in proof (2013 November 9)

The Kepler data for Q13 of V1504 Cyg and V344 Lyr were released
to public on October 22, 2013. By taking into account these new
Kepler data, we replaced the original figures 1--4 and
tables 1--2 to new ones.

\medskip

We thank the Kepler Mission team and the data calibration engineers for
making Kepler data available to the public.  We are grateful to
Enrique de Miguel for providing BK Lyn data.
This work was supported by the Grant-in-Aid
``Initiative for High-Dimensional Data-Driven Science through Deepening
of Sparse Modeling'' from the Ministry of Education, Culture, Sports, 
Science and Technology (MEXT) of Japan.  We are grateful to the referee, 
Dr. John Cannizzo, for his critical reading and constructive comments.


\begin{thebibliography}{}

\bibitem[{Balbus}, {Hawley}(1998)]{bal98ADreview}
  {Balbus}, S.~A., \& {Hawley}, J.~F.\ 1998, Reviews\ of\ Modern\ Phys., 70, 1

\bibitem[Cannizzo et~al.(1988)]{can88outburst}
  Cannizzo, J.~K., Shafter, A.~W., \& Wheeler, J.~C.\ 1988, ApJ, 333, 227

\bibitem[{Cannizzo} et~al.(2012)]{can12v344lyr}
  {Cannizzo}, J.~K., {Smale}, A.~P., {Wood}, M.~A., {Still}, M.~D., \&
  {Howell}, S.~B.\ 2012, ApJ, 747, 117

\bibitem[{Cannizzo} et~al.(2010)]{can10v344lyr}
  {Cannizzo}, J.~K., {Still}, M.~D., {Howell}, S.~B., {Wood}, M.~A., \&
  {Smale}, A.~P.\ 2010, ApJ, 725, 1393

\bibitem[Cannizzo, Wheeler(1984)]{can84ADvertical}
  Cannizzo, J.~K., \& Wheeler, J.~C.\ 1984, ApJS, 55, 367

\bibitem[{de Miguel} et~al.(2012)]{dem12erumasass}
  {de Miguel}, E., {et~al.}\ 2012, in Proc. 31st Annu. Conf., Symp. on
  Telescope Science, ed. B.~D. {Warner}, \& {et~al.} (Rancho Cucamonga, CA:
  Society for Astronomical Sciences), p.~79

\bibitem[Harvey et~al.(1995)]{har95v503cyg}
  Harvey, D., Skillman, D.~R., Patterson, J., \& Ringwald, F.~A.\ 1995, PASP,
  107, 551

\bibitem[Hellier(2001)]{hel01book}
  Hellier, C.\ 2001, Cataclysmic Variable Stars: How and why they vary (Berlin:
  Springer)

\bibitem[{Ichikawa} et~al.(1993)]{ich93SHmasstransferburst}
  {Ichikawa}, S., {Hirose}, M., \& {Osaki}, Y.\ 1993, PASJ, 45, 243

\bibitem[{Kato} et~al.(2013)]{Pdot4}
  {Kato}, T., {et~al.}\ 2013, PASJ, 65, 23

\bibitem[{Kato} et~al.(2014)]{Pdot5}
  {Kato}, T., {et~al.}\ 2014, PASJ, in press (arXiv/1310.7069)

\bibitem[{Kato} et~al.(2009)]{Pdot}
  {Kato}, T., {et~al.}\ 2009, PASJ, 61, S395

\bibitem[{Kato}, {Kunjaya}(1995)]{kat95eruma}
  {Kato}, T., \& {Kunjaya}, C.\ 1995, PASJ, 47, 163

\bibitem[{Kato}, {Maehara}(2013)]{kat13j1924}
  {Kato}, T., \& {Maehara}, H.\ 2013, PASJ, 65, 76

\bibitem[{Kato} et~al.(2012)]{Pdot3}
  {Kato}, T., {et~al.}\ 2012, PASJ, 64, 21

\bibitem[{Kato} et~al.(2010)]{Pdot2}
  {Kato}, T., {et~al.}\ 2010, PASJ, 62, 1525

\bibitem[{Kato}, {Osaki}(2013)]{kat13j1922}
  {Kato}, T., \& {Osaki}, Y.\ 2013, PASJ, 65, L13

\bibitem[{Kato}, {Uemura}(2012)]{kat12perlasso}
  {Kato}, T., \& {Uemura}, M.\ 2012, PASJ, 64, 122

\bibitem[{Larwood}(1998)]{lar98XBprecession}
  {Larwood}, J.\ 1998, MNRAS, 299, L32

\bibitem[Lin et~al.(1985)]{lin85CValphadisk}
  Lin, D. N.~C., Faulkner, J., \& Papaloizou, J.\ 1985, MNRAS, 212, 105

\bibitem[{Lubow}(1994)]{lub94impact}
  {Lubow}, S.~H.\ 1994, ApJ, 432, 224

\bibitem[{Menou}(2000)]{men00ADviscosity}
  {Menou}, K.\ 2000, Science, 288, 2022

\bibitem[{Meyer-Hofmeister}(1987)]{mey87thermal}
  {Meyer-Hofmeister}, E.\ 1987, A\&A, 175, 113

\bibitem[Mineshige(1988)]{min88uvdelay}
  Mineshige, S.\ 1988, A\&A, 190, 72

\bibitem[Mineshige, Osaki(1985)]{min85DNDI}
  Mineshige, S., \& Osaki, Y.\ 1985, PASJ, 37, 1

\bibitem[{Montgomery}(2009)]{mon09diskprecession}
  {Montgomery}, M.~M.\ 2009, ApJ, 705, 603

\bibitem[{Ohshima} et~al.(2012)]{ohs12eruma}
  {Ohshima}, T., {et~al.}\ 2012, PASJ, 64, L3

\bibitem[{Ohshima} et~al.(2011)]{ohs11qzvir}
  {Ohshima}, T., {et~al.}\ 2011, PASJ, submitted

\bibitem[{Osaki}(1989)]{osa89suuma}
  {Osaki}, Y.\ 1989, PASJ, 41, 1005

\bibitem[{Osaki}(1996)]{osa96review}
  {Osaki}, Y.\ 1996, PASP, 108, 39

\bibitem[{Osaki}, {Kato}(2013a)]{osa13v1504cygKepler}
  {Osaki}, Y., \& {Kato}, T.\ 2013a, PASJ, 65, 50

\bibitem[{Osaki}, {Kato}(2013b)]{osa13v344lyrv1504cyg}
  {Osaki}, Y., \& {Kato}, T.\ 2013b, PASJ, 65, 95

\bibitem[{Osaki}, {Meyer}(2003)]{osa03DNoutburst}
  {Osaki}, Y., \& {Meyer}, F.\ 2003, A\&A, 401, 325

\bibitem[{Patterson} et~al.(2013)]{pat13bklyn}
  {Patterson}, J., {et~al.}\ 2013, MNRAS, 434, 1902

\bibitem[{Pavlenko} et~al.(2012)]{pav12v503cyg}
  {Pavlenko}, E.~P., {Samsonov}, D.~A., {Antonyuk}, O.~I., {Andreev}, M.~V.,
  {Baklanov}, A.~V., \& {Sosnovskij}, A.~A.\ 2012, Astrophysics, 55, 494

\bibitem[Robertson et~al.(1995)]{rob95eruma}
  Robertson, J.~W., Honeycutt, R.~K., \& Turner, G.~W.\ 1995, PASP, 107, 443

\bibitem[{Skillman}, {Patterson}(1993)]{ski93bklyn}
  {Skillman}, D.~R., \& {Patterson}, J.\ 1993, ApJ, 417, 298

\bibitem[Smak(1984)]{sma84ugemdiskradius}
  Smak, J.\ 1984, Acta\ Astron., 34, 93

\bibitem[{Smak}(1985)]{sma85vwhyi}
  {Smak}, J.\ 1985, Acta\ Astron., 35, 357

\bibitem[{Smak}(2008)]{sma08zcha}
  {Smak}, J.\ 2008, Acta\ Astron., 58, 55

\bibitem[{Smak}(2012)]{sma12zchaoycaroverflow}
  {Smak}, J.\ 2012, Acta\ Astron., 62, 213

\bibitem[{Smak}(2013)]{sma13negSH}
  {Smak}, J.\ 2013, Acta\ Astron., 63, 109

\bibitem[Steeghs et~al.(2003)]{ste03iyumaSTJ}
  Steeghs, D., Perryman, M. A.~C., Reynolds, A., de Bruijne, J. H.~J., Marsh,
  T., Dhillon, V.~S., \& Peacock, A.\ 2003, MNRAS, 339, 810

\bibitem[Stellingwerf(1978)]{PDM}
  Stellingwerf, R.~F.\ 1978, ApJ, 224, 953

\bibitem[{Truss}(2005)]{tru05masstransfer}
  {Truss}, M.~R.\ 2005, MNRAS, 356, 1471

\bibitem[Warner(1995)]{war95book}
  Warner, B.\ 1995, Cataclysmic Variable Stars (Cambridge: Cambridge University
  Press)

\bibitem[{Wood} et~al.(2011)]{woo11v344lyr}
  {Wood}, M.~A., {Still}, M.~D., {Howell}, S.~B., {Cannizzo}, J.~K., \&
  {Smale}, A.~P.\ 2011, ApJ, 741, 105

\end{thebibliography}
\end{document}